\documentclass[dvips]{acta}

\usepackage{supertabular,lscape,epsfig}
\usepackage{amssymb}
\usepackage{amsmath}

\SetPages{0}{0}

\SetVol{61}{2011}

\usepackage{times}
\def\refd{\noindent\hangafter=1\hangindent=1cm}

\begin{document}
\centerline{\large\bf A CCD Search for Variable Stars of Spectral Type B}
\centerline{\large\bf in the Northern Hemisphere Open Clusters.}
\centerline{\large\bf VIII. NGC\,6834}

\vspace{0.3cm}
\centerline{by}

\vspace{0.3cm}
\centerline{M.~J e r z y k i e w i c z$^1$,\quad G.~K o p a c k i$^1$,\quad 
A.~P i g u l s k i$^1$,} 
\centerline{Z.~K o {\l } a c z k o w s k i$^1$\quad and\quad S.-L.~K i m$^2$}

\vspace{0.3cm}
\centerline{$^1$ Astronomical Institute, University of  Wroc{\l }aw, Kopernika 11,}
\centerline{51-622 Wroc{\l }aw, Poland,}
\centerline{e-mail: (mjerz,kopacki,pigulski,kolaczkowski)@astro.uni.wroc.pl}
\vspace{0.1cm}
\centerline{$^2$ Korea Astronomy and Space Science Institute, Daejeon 305-348, Korea,} 
\centerline{e-mail: slkim@kasi.re.kr}

\vspace{0.5cm}
\centerline{\it Received ...}

\vspace{1cm}
\centerline{ABSTRACT}

\vspace{0.5cm}
{\small
We present results of a CCD variability search in the field of the young open cluster 
NGC\,6834. We discover 15 stars to be variable in light. The brightest, a 
multiperiodic $\gamma$ Doradus-type variable is a foreground star. The eight fainter 
ones, including a $\gamma$ Cassiopeiae-type variable, two $\lambda$ Eridani-type 
variables, an ellipsoidal variable, an EB-type eclipsing binary, and three variable 
stars we could not classify, all have $E(B-V)$ within proper range, thus fulfilling 
the necessary condition to be members. One of the three unclassified variables may be 
a non-member on account of its large angular distance from the center of the cluster. 
Four of the six faintest variable stars, which include two eclipsing binaries and two 
very red stars showing year-to-year variations, are certain non-members. One of the 
remaining two faintest variable stars, an EA-type eclipsing binary may be a member, 
while the faintest one, a W Ursae Majoris-type variable, is probably a non-member. 

For 6937 stars we provide the $V$ magnitudes and $V-I_{\rm C}$ color indices on the 
standard system. Because of nonuniform reddening over the cluster's face, a direct 
comparison of these data with theoretical isochrones is not possible. We therefore 
obtain $E(B-V)$ from available $UBV$ photometry, determine the lower and upper bound 
of $E(B-V)$ for NGC\,6834, and then fit properly reddened Padova isochrones to the 
data. Assuming HDE\,332843, an early-F supergiant, to be a member we get $\log ({\rm 
age/yr}) ={}$7.70, $V_0 - M_{\rm V} ={}$12.10~mag. 

For 103 brightest stars in our field we obtained the $\alpha$ index, a measure of the 
equivalent widths of the H$\alpha$ line. We find H$\alpha$ emission in five stars, 
including the $\gamma$ Cas-type variable and the two $\lambda$ Eri-type variables.

\vspace{0.2cm}
}

\vspace{0.5cm}
\centerline{\bf 1. Introduction}

\vspace{0.5cm} 
This is the eighth paper in the series containing results of our search for B-type 
variable stars in young open clusters of the Northern Hemisphere.  In the preceding 
papers we presented results for the following open clusters:  NGC\,7128 (Jerzykiewicz 
{\it et al.\/}~1996, hereafter Paper I), NGC\,7235 (Pigulski {\it et al.\/}~1997, 
Paper II), NGC\,6823 (Pigulski {\it et al.\/}~2000, Paper III), NGC\,663 (Pigulski 
{\it et al.\/}~2001, Paper IV), NGC\,2169 (Jerzykiewicz {\it et al.\/}~2003, Paper V), 
NGC\,6910 (Ko{\l }aczkowski {\it et al.\/}~2004, Paper VI), and NGC\,1502 (Michalska 
{\it et al.\/}~2009, Paper VII). Altogether we discovered some 80 variable stars.  

In the present paper, we report results for NGC\,6834 (C\,1950+292, OCl\,134), a young 
open cluster in Cygnus. According to Ruprecht (1966), the Trumpler type is II\,2m, but 
Trumpler (1930) gives I\,2m. From $UBV$ photoelectric and photographic photometry of 
Hoag {\it et al.\/}~(1961), Johnson {\it et al.\/}~(1961) found the reddening to vary 
over the face of the cluster, estimated the mean color excess, $E(B-V)$, to be equal 
to 0.72~mag and derived an extinction-corrected distance modulus, $V_0 - M_{\rm V}$, 
of 12.4$\,\pm\,$0.3 (p.e.) mag. Using the same data, Becker (1963) and Hoag and 
Applequist (1965) reduced the latter value to 11.7~mag and 11.9~mag, respectively. 
From $UBV$ photographic photometry, F\"unfschilling (1967) obtained a still smaller 
value of 11.6~mag; curiously, his $E(B-V)$ was equal to 0.64~mag. Another $UBV$ 
photographic study (Moffat 1972, Moffat and Vogt 1974) resulted in $E(B-V) 
={}$0.72~mag, $V_0 - M_{\rm V} ={}$11.65~mag and an age of 8$\times$10$^7$~yr. 

Paunzen {\it et al.\/}~(2006) included NGC\,6834 in a search for peculiar stars in 
open clusters. From CCD observations in the $g_1 g_2 y$ photometric system, they found 
ten candidates in the cluster, including two questionable ones. In addition, Paunzen 
{\it et al.\/}~(2006) derived $E(B-V) ={}$0.70$\,\pm\,$0.05~mag, $V_0 - M_{\rm V} 
={}$11.4$\,\pm\,$0.4~mag and an age of (7.9$\,\pm\,$2.1)$\times$10$^7$~yr. The 
distance obtained by these authors, 1930$\,\pm\,$32~pc, is virtually identical with 
that given by Trumpler (1930) in his table 16.  

The ``Catalogue of stars in the Northern Milky Way having H$\alpha$ in emission'' of 
Kohoutek and Wehmeyer (1999), based on the data published throughout 1994, lists three 
NGC\,6834 stars. More recently, Miller {\em et al.\/}~(1996) reported finding six 
Be-star candidates, showing enhanced H$\alpha$(continuum)/H$\alpha$ index. In 
addition, from a $BV$ color-magnitude diagram and Geneva isochrones these authors 
derived an age of about 5$\times$10$^7$~yr, a mean reddening, $E(B-V) \approx$ 
0.7~mag, and $V_0 - M_{\rm V} ={}$12.2~mag. NGC\,6834 was also included by Mathew {\it 
et al.\/}~(2008) in an extensive search for H$\alpha$ emission-line stars in young 
open clusters, carried out on low-resolution CCD slitless grism spectrograms. Four Be 
stars were found in the cluster. However, Mathew {\it et al.\/}~(2008) warn that their 
slitless spectroscopy technique identifies only stars with emission in H$\alpha$ above 
the continuum, so that the numbers of Be stars they give should be taken as lower 
limits. In the same paper, the cluster's age estimate is revised to 
4$\times$10$^7$~yr. 

In the next section we describe our observations and reductions. Section 3 contains a 
detailed discussion of the stars we discovered to be variable in light. The range of 
$E(B-V)$ for NGC\,6834 is determined in Section 4.1, while the color-magnitude diagrams 
are presented and fitted with theoretical isochrones in Section 4.2. The $\alpha$-index 
measurements are reported in Section 5. In the last section we discuss and summarize 
our results.

Throughout the paper we use the WEBDA\footnote{http://www.univie.ac.at/webda/} 
numbering system, preceding the WEBDA numbers with a W. In most cases, 
the stars from W1 to W205 are identical with those having the same numbers in 
F\"unfschilling (1967). Exceptions will be noted in Section 3 and 4.1. 
 
\vspace{0.5cm}
\centerline{\bf 2. Observations and Reductions}

\vspace{0.5cm}
We have observed NGC\,6834 from two sites, the Bia{\l }k\'ow Observatory of the Wroc{\l 
}aw University, Poland, and the Bohyunsan Optical Astronomy Observatory (BOAO), South 
Korea. At Bia{\l }k\'ow, the observations were obtained on a number of nights in 2001, 
and on a single night in 2010. In 2001, we used equipment described in Papers I and II, 
viz., a 60-cm reflecting telescope with a 576$\times$384 pixels CCD camera, mounted in 
the Cassegrain focus, an autoguider, a set of $BV(RI)_{\rm C}$ filters, and a pair 
H$\alpha$ interference-filters, narrow (having FWHM equal to 3 nm) and wide (FWHM equal 
to 20 nm). Time-series observations were carried out on 27 nights in the interval from 
April 28 to October 12, 2001. However, the bulk of the data was obtained on 25 nights 
between June 25 and August 26. On 24 nights the time-series observations were obtained 
with two filters, $V$ and $I_{\rm C}\/$, on the remaining three nights, with the $I_{\rm 
C}$ filter. Altogether, some 880 frames were taken through the $V$ filter, and some 1100 
frames, through the $I_{\rm C}$ filter. The integration time ranged from 120~s to 260~s 
for the $V$ frames, and from 90~s to 200~s for the $I_{\rm C}$ frames, but a few frames 
were taken with the integration time of 500~s and 300~s for $V$ and $I_{\rm C}$, 
respectively. The H$\alpha$ measurements were obtained on two nights, October 12/13 and 
14/15, 2001 for 103 stars brighter than 16.8 magnitude in $V\/$. For each star, six frames 
were taken through the wide H$\alpha$ filter, and six, through the narrow H$\alpha$ 
filter. In order to improve the signal to noise ratio, the frames were co-added, 
resulting in one wide H$\alpha$ filter frame with a total integration time of 2188~s, 
and one narrow H$\alpha$ filter frame with a total integration time of 13466~s. All 
$V$, $I_{\rm C}$, and H$\alpha$ frames were calibrated in the same way as in Paper I and 
then reduced by means of Stetson's (1987) DAOPHOT II package. 

On the single night in 2010, July 10/11, the observations were taken for the purpose 
of standardizing our photometry. We used the same 60-cm telescope, but a new CCD 
camera, Andor DW432-BV (see Paper VII). On this night, we observed 
NGC\,6834 together with a nearby open cluster NGC\,6823, the $VI_{\rm C}$ photometry 
of which had been tied to the Landolt's (1992) standards (see Paper III). 

At BOAO, the equipment consisted of a 1.8-m Cassegrain telescope and a 2048$\times$2048 
pixels CCD camera. On two nights in 2001, June 15 and 17, a $V$-filter time-series 
observations were carried out. There were 57 and 14 frames taken on the first and second 
night, respectively. Two-color observations were obtained on June 16, 2002. On this 
night, 14 frames were taken in $V$ and four, in $I_{\rm C}$. The integration time ranged 
from 30~s to  400~s for the $V$ frames, and was equal to 60~s for the $I_{\rm C}$ frames. 
The calibrations and reductions were carried out by means of the same procedure as that 
used for the Bia{\l }k\'ow frames.  

%
\begin{figure}[htb]
\hbox to\hsize{\hss\includegraphics{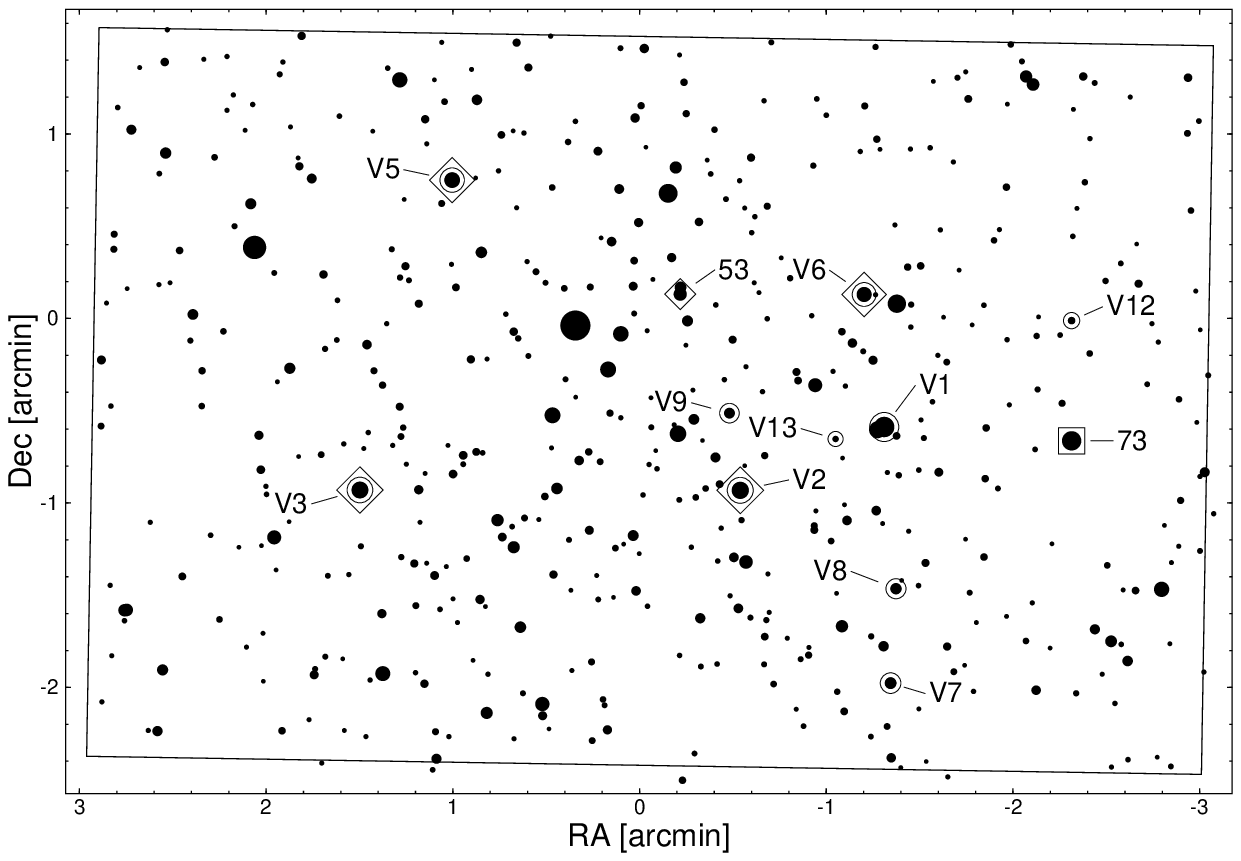}\hss} 
\FigCap{The field of NGC\,6834 observed from Bia{\l }k\'ow. North is up, East to the 
left. RA and Dec are given relative to 19$^{\rm h}$52$^{\rm m}$12$^{\rm s}$ and 
29$^{\rm o}$24$^\prime$30$^{\prime\prime}$, epoch 2000.0. The stars we discovered to 
be variable in light are shown with points in open circles, and those to have 
emission at H$\alpha$, with symbols in open diamonds. The variable stars are labeled 
V1, V2, etc. The H$\alpha$-emission star W53, which we did not find to vary, is 
labeled 53. For W73 (the point in an open square labeled 73), the Be star listed 
in the catalogue of Kohoutek and Wehmeyer (1999), we found no emission at H$\alpha$ 
(see Fig.\ 11).} 
\end{figure}

Fig.~1 shows the 6$^\prime \times$4$^\prime$ Bia{\l }k\'ow field of NGC\,6834. The stars 
we found to be variable in light, and those to have emission at H$\alpha$ are indicated. 
The 11.5$^\prime \times$12$^\prime$ field observed from BOAO is shown in Fig.~2. The 
stars outside the Bia{\l }k\'ow field discovered to be variable in light and the Be star 
W106 (see Section 5) are indicated. 

%
\begin{figure}[htb] 
\hbox to\hsize{\hss\includegraphics{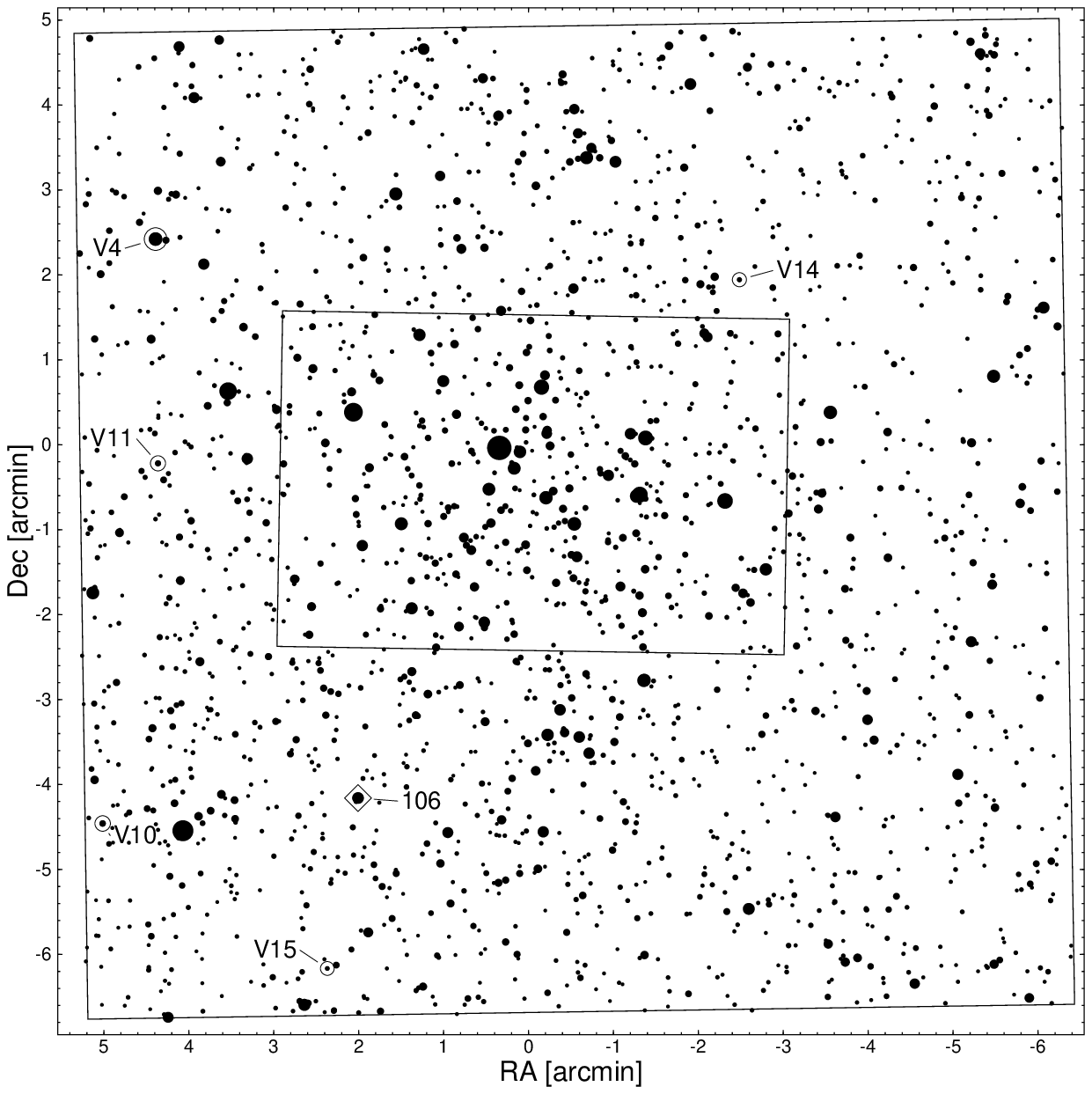}\hss}  
\FigCap{The field of NGC\,6834 observed from BOAO; the rectangle in the middle 
delineates the field observed from Bia{\l }k\'ow (see Fig.\ 1). North is up, East to 
the left. RA and Dec are given relative to 19$^{\rm h}$52$^{\rm m}$12$^{\rm s}$ and 
29$^{\rm o}$24$^\prime$30$^{\prime\prime}$, epoch 2000.0. The stars outside the 
Bia{\l }k\'ow field discovered to be variable in light are shown with points in open 
circles and labeled. The point in an open diamond is the Be star W106 (see 
Section 5).} 
\end{figure}

Following Kopacki {\it et al.\/}~(2008), we defined our instrumental $V$- and $I_{\rm 
C}$-filter magnitudes, to be referred to below as $v$ and $i$, by means of DAOPHOT 
magnitudes of a set of bright, unsaturated stars in a reference frame. In the 
remaining frames, the DAOPHOT magnitudes of these stars were used to compute the mean 
magnitude difference between a given frame and the reference frame. Adding this number 
to the DAOPHOT magnitudes of all stars in the frame in question yielded the 
instrumental magnitudes, $v$ and $i$. Note that the instrumental magnitudes derived in 
this way are corrected for the first-order (i.e., color independent) atmospheric 
extinction. 

The transformation of our instrumental $v$ and $i$ magnitudes to the standard system was 
carried out using the July 10/11, 2010 observations. To begin with, we derived aperture 
magnitudes for the brightest stars in both fields, NGC\,6834 and NGC\,6823, by means of 
the DAOPHOT procedure DAOGROW (Stetson 1990). Then, we corrected these magnitudes for the 
effect of first-order atmospheric extinction by fitting a linear extinction equation 
simultaneously to the magnitudes in both fields, separately for the $V$- and $I_{\rm 
C}$-filter magnitudes. Since the observations of both clusters were obtained at similar 
average air masses, the extinction-corrected magnitudes were derived for the average 
value of the air mass, $\langle X\rangle={}$1.136. In the next step, we determined the 
following transformation equations using 34 stars from NGC\,6823:

\vskip0.7\baselineskip
 \hbox to\hsize{\hfill\vbox{\tabskip=1pt 
  \halign{%
   \hfil#\tabskip=5.5pt&%
   #\hfil\tabskip=0.5cm&%
   #\hfil\tabskip=1pt\cr
   $V-v'$&         $={}-0.043\,(v'-i')-5.066$,& $\sigma={}$0.007 mag,\cr
   $V-I_{\rm C}$& $={}+0.932\,(v'-i')+0.554$,& $\sigma={}$0.010 mag,\cr
  }}\hfill}
\vskip0.4\baselineskip

\noindent
where uppercase letters denote standard magnitudes, and $v'$ and $i'$, the 
extinction-corrected magnitudes; $\sigma$ is the standard deviation of the fit. The 
$v'$ and $i'$ magnitudes should not be confused with the $v$ and $i$ magnitudes 
defined in the preceding paragraph. 

The above equations were applied to convert the NGC\,6834 $v'$ and $i'$ magnitudes to 
the standard system. The final step consisted in using these standardized magnitudes, 
$V$ and $I_{\rm C}$, for deriving overall transformations of the average instrumental 
magnitudes, $\langle v\rangle$ and $\langle i\rangle$, from both NGC\,6834 data sets, 
Bia\l{}k\'ow and BOAO, to the standard system. We obtained the following 
transformation equations:

\vskip0.7\baselineskip
 \hbox to\hsize{\hfill\vbox{\tabskip=1pt 
  \halign{%
   \hfil#\tabskip=5.5pt&%
   #\hfil\tabskip=0.5cm&%
   #\hfil\tabskip=0.5cm&%
   #\hfil\tabskip=1pt\cr
   $V-\langle v\rangle$&  $={}+0.090\,(\langle v\rangle-\langle i\rangle)+0.422$,& 
                                                               $\sigma={}$0.016 mag,& $N={}$51,\cr
   $V-I_{\rm C}$&         $={}+1.023\,(\langle v\rangle-\langle i\rangle)+0.862$,& 
                                                                       $\sigma={}$0.009 mag,&\cr
  }}\hfill}
\vskip0.4\baselineskip

\noindent
for Bia\l{}k\'ow, and

\vskip0.7\baselineskip
 \hbox to\hsize{\hfill\vbox{\tabskip=1pt 
  \halign{%
   \hfil#\tabskip=5.5pt&%
   #\hfil\tabskip=0.5cm&%
   #\hfil\tabskip=0.5cm&%
   #\hfil\tabskip=1pt\cr
   $V-\langle v\rangle$&  $={}-0.019\,(\langle v\rangle-\langle i\rangle)+2.239$,& 
                                                               $\sigma={}$0.010 mag,& $N={}$149,\cr
   $V-I_{\rm C}$&$={}+0.914\,(\langle v\rangle-\langle i\rangle)+1.219$,& $\sigma={}$0.012 mag,&\cr
  }}\hfill}
\vskip0.4\baselineskip

\noindent
for BOAO; $N$ is the number of stars used in the least-squares solutions.

%
\begin{figure}[htb] 
\hbox to\hsize{\hss\includegraphics{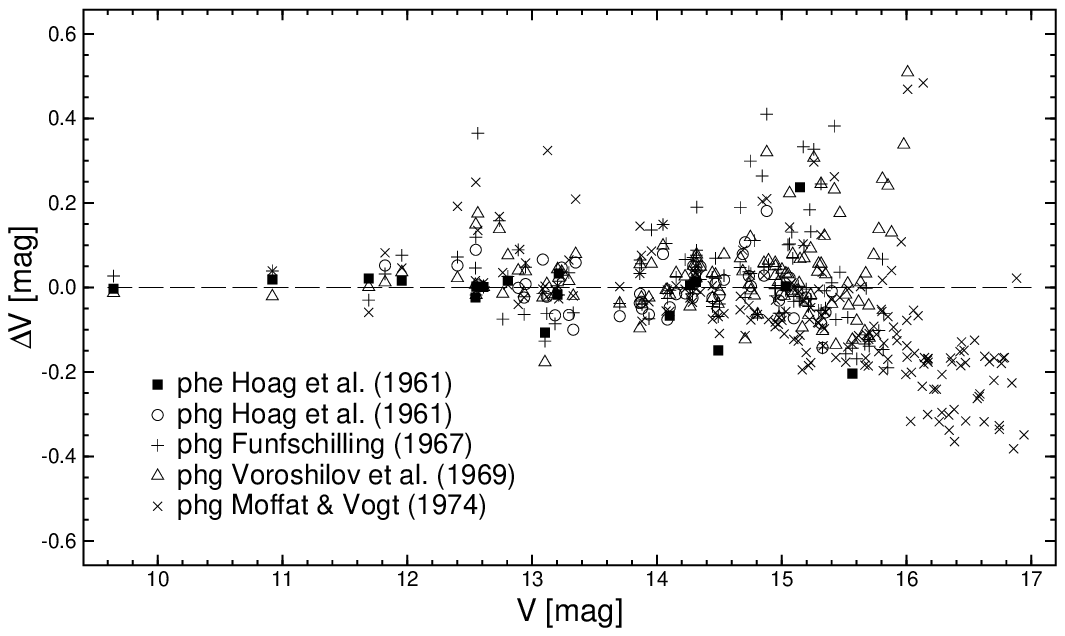}\hss}  
\FigCap{A comparison of the $V$ magnitudes derived in this paper with those from the 
photoelectric and photographic photometry of Hoag {\em et al.\/}~(1961) (filled 
squares and open circles, respectively) and from the photographic data of 
F\"unfschilling (1967) (plus signs), Voroshilov {\em et al.\/}~(1969) (triangles), and 
Moffat and Vogt (1974) (crosses). $\Delta V$ is the difference between the Bia{\l 
}k\'ow $V$ magnitude and that given by others. The dashed line indicates zero 
ordinate.} 
\end{figure}

Fig.~3 shows the differences between the Bia{\l }k\'ow $V$ magnitudes and those from 
Hoag {\em et al.\/}~(1961), F\"unfschilling (1967), Voroshilov {\em et al.\/}~(1969), 
and Moffat and Vogt (1974); Table 1 gives the mean values of these differences, 
$\langle\Delta V\rangle$, standard deviations, $\sigma$, and the number of data used, 
$N$. The mean differences were determined for $V<$15.2~mag using a $3\sigma$-clipping 
algorithm. The latter eliminated the differences obtained from images of stars that 
were blended in earlier photometry but resolved on our CCD frames and several cases of 
misidentification. We shall return to this point in the next section and in Section 
4.1. 

As can be seen from Fig.~3 and Table 1, the agreement between our $V$ magnitudes and 
those available in the literature is very good for $V<$15.6~mag. Unfortunately, we 
cannot comment on the reliability of our $(V-I_{\rm C})$ color indices, because of the 
lack of any other photometric data for NGC\,6834 in the $I_{\rm C}$ passband.

\vskip1cm
\def\mp{\omit\hfil--\hfil}
\vbox{\noindent\hbox to\hsize{\hss T a b l e \quad 1\hss}\par
\vskip0.3\baselineskip\small
\noindent\hfil Mean differences between our and other $V$ magnitudes\par
\vskip\baselineskip
\normalbaselineskip=\baselineskip
\setbox\strutbox=\hbox{\vrule height0.7\normalbaselineskip%
 depth0.3\normalbaselineskip width0pt}
 \hbox to\hsize{\hfill\vbox{\tabskip=6pt 
  \halign{%
   #\hfil\tabskip=20pt&
   \hfil#&
   \hfil#&
   \hfil#\hfil&
   \hfil#\hfil
   \tabskip=6pt\cr
   \noalign{\hrule\vskip3pt}
   Photometry& Type& $\langle\Delta V\rangle$& $\sigma$& $N$\cr
   \noalign{\vskip1pt}
      && [mag]& [mag]& \cr
   \noalign{\vskip3pt\hrule\vskip3pt}
   Hoag {\em et al.\/}~(1961)& phe&     0.006&    0.016&   13\cr
   Hoag {\em et al.\/}~(1961)& phg&  $-$0.001&    0.050&   43\cr
   F\"unfschilling (1967)&       phg&     0.023&    0.066&   85\cr
   Voroshilov {\em et al.\/}~(1969)&          phg&     0.017&    0.045&   78\cr
   Moffat and Vogt (1974)&     phg&  $-$0.001&    0.083&   86\cr
  \noalign{\vskip3pt\hrule}
  }}\hfill}}
\vskip1cm

Tables A1 and A2, containing the equatorial coordinates, RA and Dec, and the mean $V$ 
and $I_{\rm C}$ magnitudes of all stars we detected in the Bia{\l }k\'ow and BOAO 
fields of NGC\,6834, respectively, can be downloaded from the {\em Acta Astronomica 
Archive\/} (see the cover page). In the tables, the variable stars are identified with 
the labels V1, V2, etc. In Table A1,  the $\alpha$ index is listed for the 103 
brightest stars of the Bia{\l }k\'ow field. Tables containing the $V$ and $I_{\rm C}$ 
time-series observations of the stars we found to be variable in light can be also 
downloaded from the {\em Archive\/}. In addition to the HJD epochs of the middle of 
the exposure and the $V$ and $I_{\rm C}$ magnitudes, these tables include the DAOPHOT 
formal errors of the magnitudes, $e_V$ and $e_I$. 

\vspace{0.5cm}
\centerline{\bf 3. Variable Stars}

\vspace{0.5cm} 
We have analyzed for variability the $V$ and $I_{\rm C}$ magnitudes of all stars in the 
Bia{\l }k\'ow field with more than 100 $I_{\rm C}$-filter observations. The analysis 
included inspecting light-curves by eye and computing least-squares (LS) power spectra 
in the frequency range from 0 to 30~d$^{-1}$. Here and in what follows, by `power' we 
mean $1 - \sigma^2(f)/\sigma^2$, where $\sigma^2(f)$ is the variance of a least-squares 
fit of a sine curve of frequency $f$ to the data, and $\sigma^2$ is the variance of the 
same data. If a frequency was identified from the power spectrum, the data were 
prewhitened with a sine-curve of this frequency and then the residuals were analyzed. If 
a second frequency was found, the data were prewhitened with the two frequencies, etc. 
In computing the LS spectra and prewhitening, we applied weights inversely proportional 
to the squares of the formal errors of the magnitudes, $e_V^2$ or $e_I^2$, i.e., we 
multiplied each equation of condition by a square root of the weight. 

On the average, the formal errors of the BOAO magnitudes were a factor of $\sim$3 
smaller than the Bia{\l }k\'ow ones for the same stars. Thus, the BOAO magnitudes would 
have much higher weights than the Bia{\l }k\'ow magnitudes obtained on nights of 
comparable photometric quality. In order to compensate for this, we multiplied the 
formal errors of the BOAO magnitudes by 3. 

The stars outside the Bia{\l }k\'ow field were examined for variability by inspecting 
the $V$ magnitudes obtained at BOAO on the three nights mentioned in Section 2, June 15 
and 17, 2001, and June 16, 2002. 

We shall now discuss the stars that we found to be variable in light. Each star is first 
referred to by the label V1, V2, etc. If a star appears in WEBDA, its WEBDA number is 
also given. 

{\bf V1 = W72.} In this case the WEBDA number is not identical with that of 
F\"unfschilling (1967). The latter author notes that his number 72 is a blend. On our 
CCD frames, the blend is resolved into two stars. The brighter one is V1 = W72, while 
the fainter one is W2098. In the table of photoelectric data of Hoag {\it et 
al.\/}~(1961), F\"unfschilling's (1967) number 72 is number 5. The difference of the 
mean $V$ magnitude of V1 = W72 from Table A1 and the $V$ magnitude of number 5 from 
Hoag {\it et al.\/}~(1961) amounts to 0.45~mag, and therefore was eliminated from 
Fig.\ 3 by the $3\sigma$-clipping algorithm. If, however, V1 = W72 were replaced by 
its blend with W2098, the difference would be reduced to $-$0.006~mag, in a 
one-$\sigma$ agreement with $\langle\Delta V\rangle$ listed in Table 1. 

As can be seen from the {\em top panels\/} of Fig.\ 4, the LS power spectra of the $V$ 
and $I_{\rm C}$ magnitudes of V1 show the highest peaks at different frequencies. 
Denoting these frequencies in the order of decreasing amplitudes in the three-frequency 
LS fits (see below), we have $f_2 ={}$1.668~d$^{-1}$ and $f_1 ={}$1.532~d$^{-1}$. The 
spectra computed after prewhitening the $V$ magnitudes with $f_2$ and the $I_{\rm C}$ 
magnitudes with $f_1$ showed the highest peaks in reversed order (see the {\em 
upper-middle panels\/} of Fig.\ 4). In the next step, we computed the power spectra of 
the data  prewhitened simultaneously with $f_1$ and $f_2$. Now the highest peaks appeared 
at frequencies which differed by 1~d$^{-1}$. In the case of $V,$ the frequency of the highest 
peak was 1.626~d$^{-1}$, while in the case of $I_{\rm C}$ it was 0.626~d$^{-1}$, but the alias 
peak at 1.626~d$^{-1}$ was of almost the same height. As can be seen from the {\em 
lower-middle right-hand panel\/} of Fig.\ 4, the $I_{\rm C}$ peak at 0.626~d$^{-1}$ may 
have been enhanced by low-frequency noise. We therefore conclude that the true value of 
the third frequency, $f_3$, is equal to 1.626~d$^{-1}$. 

%
\begin{figure}[htb] 
\hbox to\hsize{\hss\includegraphics{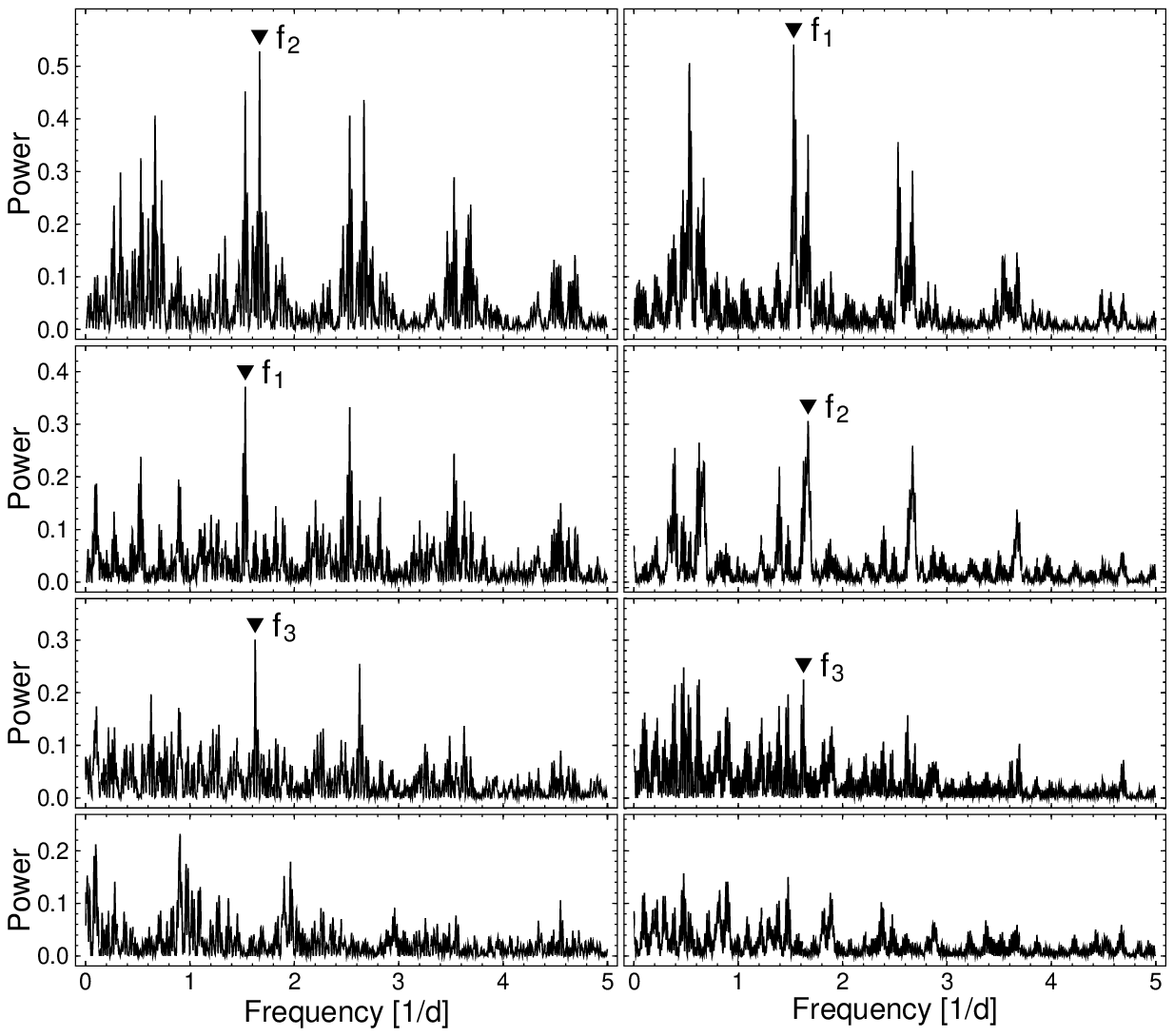}\hss} 
\FigCap{The LS power spectra of the $V$ magnitudes (the {\em left-hand panels\/}) and the 
$I_{\rm C}$ magnitudes (the {\em right-hand panels\/}) of V1. The spectra of the 
unaltered data are shown in the {\em top panels\/}, the spectrum of the $V$ magnitudes 
prewhitened with $f_2 ={}$1.668~d$^{-1}$ is shown in the {\em upper-middle left-hand 
panel\/}, that of the $I_{\rm C}$ magnitudes prewhitened with $f_1 ={}$1.532~d$^{-1}$ is 
shown in the {\em upper-middle right-hand panel\/}, the spectra computed after 
prewhitening the data with the two frequencies, $f_1$ and $f_2$, are shown in the {\em 
lower-middle panels\/}, and those computed after prewhitening with the three frequencies, 
$f_1$, $f_2$, and $f_3 ={}$1.626~d$^{-1}$, are shown in the {\em bottom panels}.} 
\end{figure}

The peculiar situation of the reversed order of $f_1$ and $f_2$ seen in the power 
spectra in the {\em top panels\/} of Fig.\ 4 is caused by the presence of $f_3$. 
Indeed, subtracting the signal of this frequency from the $V$ magnitudes results in a 
power spectrum in which the peak at $f_1$ is stronger than that at $f_2$. It is thus 
clear that in case of the unaltered $V$ data the interference of the three components 
enhances the $f_2$ peak to such a degree that it becomes stronger than $f_1$. 

In order to refine the frequencies $f_1$, $f_2$, and $f_3$, we solved the equation
\begin{equation} 
m = \langle m \rangle + \sum_{i=1}^3 A_i(m) \cos [2 \pi f_i(m) t  
+ \phi_i(m)], 
\end{equation}
where $m$ is either $V$ or $I_{\rm C}$, by means of the nonlinear least-squares method. 

The standard deviations of the solutions amounted to 5.8~mmag and 4.2~mmag for $V$ and 
$I_{\rm C}$, respectively; and the mean magnitudes were $\langle V \rangle ={}$11.599~mag 
and $\langle I_{\rm C} \rangle ={}$11.165~mag. The frequencies, $f_i$, amplitudes 
(half-ranges), $A_i(V)$ and $A_i(I_{\rm C})$, and the epochs of maximum light, HJD$_i$, 
are listed in Table 2; the epochs of maximum light are given instead of the phases, 
$\phi_i$. The frequencies and the epochs of maximum light are the weighted mean values of 
the $V$ and $I_{\rm C}$ solutions. 

\begin{table} 
\begin{center} 
\centerline{T a b l e \quad 2} 
{Parameters of the three-frequency LS fit to the light-variation of V1} 
\vspace{0.3cm} 
{\small 
\begin{tabular} 
{ccrrc} \\ 
\hline\noalign{\smallskip} 
$i$& $f_i$&          $A_i(V)$& $A_i(I_{\rm C})$& HJD$_i-{}$2452000\\ 
   & [d$^{\rm -1}$]& [mmag]&   [mmag]&           [d]\\ 
\noalign{\smallskip}\hline\noalign{\smallskip} 
1& 1.53102${}\pm{}$0.00012 & 18.9${}\pm{}$0.5& 14.2${}\pm{}$0.3& 109.7241${}\pm{}$0.0018\\
2& 1.66870${}\pm{}$0.00014 & 15.1${}\pm{}$0.5&  8.6${}\pm{}$0.4& 109.7504${}\pm{}$0.0024\\
3& 1.62494${}\pm{}$0.00023 & 10.5${}\pm{}$0.5&  6.2${}\pm{}$0.3& 110.0203${}\pm{}$0.0032\\
\noalign{\smallskip}\hline\noalign{\smallskip}                     
\end{tabular} 
 } 
\end{center} 
\end{table}

In the {\em bottom panels\/} of Fig.\ 4 there are shown the LS power spectra of the data 
prewhitened with the three frequencies. A number of peaks, especially in the {\em 
left-hand panel\/}, protrude from the noise. We believe that the corresponding 
frequencies are spurious. 

According to Turner (1976), W72 is a foreground star of MK type F2\,V. Taking 
into account the MK type, the values of the three frequencies in Table 2, and the 
wavelength dependence of the amplitudes, we conclude that V1 is a multiperiodic 
$\gamma$ Dor-type variable. 

{\bf V2 = W67.} The star is listed in the catalogue of Kohoutek and Wehmeyer 
(1999). We confirm the H$\alpha$ emission (see Fig.\ 11). As can be seen from Fig.\ 5, 
the star shows irregular light-variations on time scales from hours to months. From 
this and the H$\alpha$ emission, we conclude that V2 is a $\gamma$ Cas-type 
variable.

%
\begin{figure}[htb] 
\hbox to\hsize{\hss\includegraphics{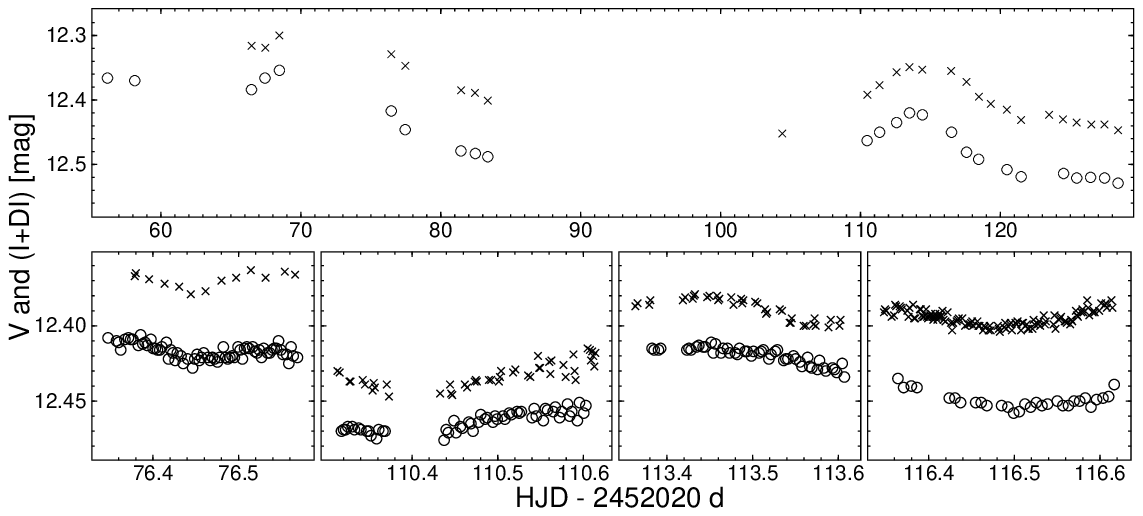}\hss} 
\FigCap{The $V$ magnitudes and the $I_{\rm C}$ magnitudes offset by DI (circles and 
crosses, respectively) of V2, plotted as a function of Heliocentric Julian Date. The 
nightly mean magnitudes are displayed in the {\em upper panel\/}. Individual 
magnitudes on four nights in the {\em lower panels\/} illustrate the short time-scale 
variation mentioned in the text. DI is equal to 0.88~mag in the {\em upper panel\/}, 
and to 0.92~mag in the {\em lower panels\/}. Note the difference in ordinate scale 
between the {\em upper\/} and {\em lower panels}.}
\end{figure}

{\bf V3 = W23.} According to Hoag and Applequist (1965), the MK type is B5\,V 
and the star is a member of the cluster. Our measurements show H$\alpha$ emission (see 
Fig.\ 11).  

Our $V$ and $I_{\rm C}$ magnitudes of V3 can be fit with either a 0.3084-d or a 
0.3065-d sine-curve. In the first case, the amplitudes amount to 5.2$\,\pm\,$0.3~mmag 
for $V$, and 7.7$\,\pm\,$0.3~mmag for $I_{\rm C}$. In the second case, the amplitudes 
are equal to 5.5$\,\pm\,$0.3 mmag for $V$, and 7.3$\,\pm\,$0.3~mmag for $I_{\rm C}$. 
In both cases, the LS power spectra of the residuals show no significant peaks. We 
conclude that the star is singly periodic. In the following discussion we shall assume 
the 0.3084-d period to be the correct one; our conclusions would remain valid if the 
other period were assumed instead. 

As its MK type attests, the star falls in the middle of the SPB-stars region of the 
\hbox{H-R} diagram. However, SPB stars of mid-B spectral type have periods a factor of 
3 to 5 longer than the periods derived above. Periods around 0.3~d, along with periods 
around 0.6~d and 8~d, have been detected in the B5\,Ve star HD\,163868 by Walker {\em 
et al.\/}~(2005). These authors consider HD\,163868 to be a prototype of a class of 
rapidly rotating SPB stars, which they designate as SPBe. Dziembowski {\em et 
al.\/}~(2007) have interpreted the variation of HD\,163868 as pulsation in $g$ modes, 
viz., $\ell ={}$2, $m ={}$2 (prograde) modes in case of the 0.3~d periods, $\ell 
={}$2, $m ={}$0 and $m ={}-$2 for the intermediate periods, and $\ell ={}$1, $m ={}-$1 
for the longest periods. The MK type and emission at H$\alpha$ make V3 similar to 
HD\,163868. While, however, the latter star is multiperiodic, V3 is not. This is the 
main objection to classifying V3 as an SPBe star. 

Another possibility, also suggested by the star's H$\alpha$ emission, is that V3 is a 
$\lambda$ Eri-type variable, i.e., a Be star with the light variation caused by 
rotational modulation. As demonstrated by Balona (1990), the photometric period in a 
variable of this type is equal to the period of rotation or its low-order submultiple. In 
case of V3, equating the 0.3084-d period with the rotation period and assuming a radius 
of 3.9~R$_\odot$, consistent with the MK type of B5\,V (Lang  1992), leads to equatorial 
velocity of rotation equal to 640~km~s$^{-1}$. This is 30 percent higher than the breakup 
velocity of a B5\,V star (Slettebak 1966). If the rotation period were assumed to be 
equal to twice the 0.3084-d period, the velocity of rotation would amount to about 65 
percent of the breakup velocity, a value which is acceptable. Thus, V3 may be a $\lambda$ 
Eri-type variable with a double-wave light curve. The scatter seen in the light curves of 
V3 in the {\em top left-hand panel\/} of Fig.\ 6 may be due in part to irregular 
variations which in $\lambda$ Eri-type variables often accompany the rotational 
modulation. 

%
\begin{figure}[htb] 
\hbox to\hsize{\hss\includegraphics{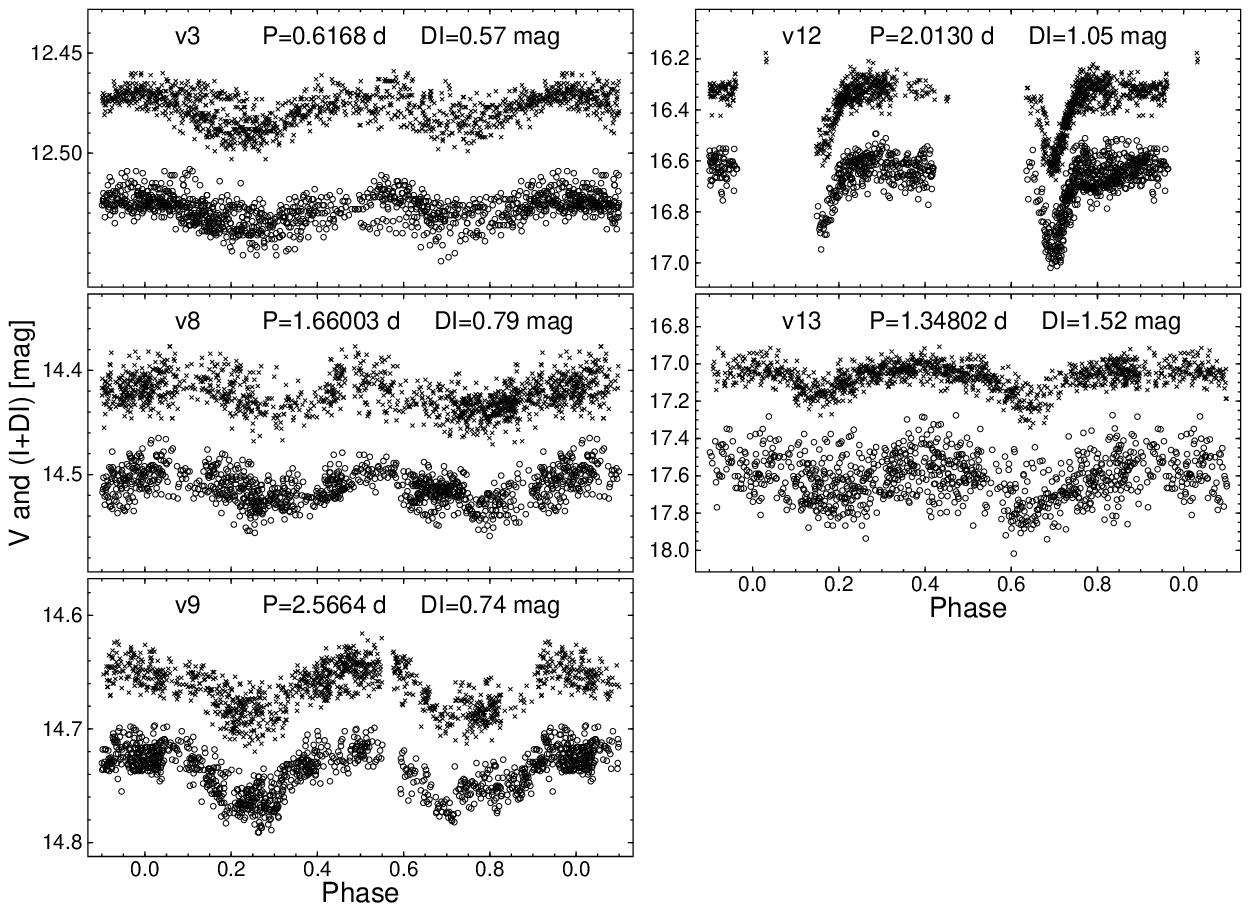}\hss}  
\FigCap{The $V$ magnitudes and the $I_{\rm C}$ magnitudes offset by DI (circles and 
crosses, respectively) of the five periodic variable stars, V3, V8, V9, V12, and V13, 
plotted as a function of phase of periods equal to $P\!\!$, twice the value derived from 
the LS spectra. The epochs of zero phase are arbitrary.} 
\end{figure}

Finally, the double-wave light variation seen in the {\em top left-hand panel\/} of 
Fig.~6 may be due to aspect changes of a tidally distorted component (or components) 
in a binary system. If this were the case, V3 would be an ellipsoidal variable (Ell in 
the notation of GCVS\footnote{http://www.sai.msu.su/groups/cluster/gcvs/gcvs/} ) with 
the orbital period equal to 0.6168~d, twice the value derived from the LS spectra. We 
shall presently discuss this possibility in some detail. 

To begin with, let us assume the secondary component to be a main-sequence star. Now, 
taking the masses and radii from Lang (1992), we find from the Kepler's third law that 
the earliest spectral type of the secondary component for which the sum of the 
component's radii does not exceed the semimajor axis of the orbit is A0. 

Let us now estimate the amplitude (half-range) of ellipsoidal light-variation of a B5\,V 
+ A0\,V system. Assuming zero eccentricity, the amplitude of the contribution of a 
component of radius $R$ to the flux variation of the system, $\delta I\!\!$, can be 
obtained from the following formula:
\begin{equation} 
\delta I = 1.5 A_\lambda q (R/a)^3 \sin^2 i, 
\end{equation} 
where $A_\lambda$ is the photometric distortion parameter of Russell and Merrill 
(1952), $q$ is the mass ratio, $a$ is the diameter of the relative orbit, and $i$ is 
the inclination of the orbit. This formula has been shown by Ruci\'nski (1969) to be 
adequate for early-type stars provided that $\delta I <$ 0.07, a condition met in the 
present case. 

After neglecting the second and higher powers of $\delta I\!\!$, Eq.\ (2) can be 
rewritten as
\begin{equation} 
\delta m = 1.629 A_\lambda q (R/a)^3 \sin^2 i, 
\end{equation}  
where $\delta m$ is the amplitude of the ellipsoidal light-variation expressed in 
magnitudes. We note that this formula is equivalent to equation (6) of Morris (1985). 

In order to compute the photometric distortion parameter, 
\begin{equation} 
A_\lambda = (15 + u_\lambda) (1+y_\lambda)/(30-10 u_\lambda), 
\end{equation} 
one needs the linear limb-darkening coefficient, $u_\lambda$, and the 
gravity-darkening coefficient, $y_\lambda$. We derived $u_\lambda$ from 
D\'{\i}az-Cordov\'es {\it et al.\/}~(1995) for $V$ and from Claret {\it et 
al.\/}~(1995) for $I_{\rm C}$, assuming the B5\,V component's $T_{\rm eff}$ to be 
equal to 15400 K, and that of the A0\,V component to be equal to 9520 K; we assumed 
$\log g ={}$4.0 for both components. The gravity-darkening coefficients were obtained 
from Ruci\'nski's (1969) table 3 where $y_\lambda$ is given as a function of spectral 
type and $\lambda$; in the case of $I_{\rm C}$ we had to extrapolate beyond 6790 {\AA}, 
the longest wavelength in the table. 

The photometric distortion parameter decreases with increasing $\lambda$. For $V$ it is 
thus greater than for $I_{\rm C}$. This signals a problem, because the observed $V$ 
amplitude is smaller than the $I_{\rm C}$ amplitude (see above and Table 3). In the 
present case of the B5\,V + A0\,V system, reproducing the observed amplitudes requires 
$i ={}$9.5$^{\rm o}$ for $V$, and $i ={}$12.5$^{\rm o}$ for $I_{\rm C}$. 

Apart from the problem with wrong wavelength-dependence of the amplitudes, the model 
leads to a (slight) $B-V$ discrepancy because the $B-V$ color-index of an unresolved 
B5\,V + A0\,V double would be 0.025~mag redder (i.e., larger) than that of a B5\,V 
star. In fact, dereddened $B-V$ color-index of V3 is consistent to within 0.01~mag  
with its MK type (see Hoag and Applequist 1965). 

This discrepancy would disappear if the secondary component's spectral type were later 
than F5, but it would have to be M0 or later in order to get $i >$ 22$^{\rm o}$. 
However, the difference between inclinations that reproduce the observed $V$ and $I_{\rm 
C}$ amplitudes increases quickly with advancing spectral type. For example, for the 
secondary's spectral type of M5, there is $i ={}$34$^{\rm o}$ for $V$, and $i 
={}$49$^{\rm o}$ for $I_{\rm C}$. An additional complication is that in the latter case 
$a \cos i$ is smaller than the sum of the component's radii, so that -- contrary to the 
observations -- the system would be eclipsing. 

We conclude that ellipsoidal variation is less satisfactory in accounting for the 
light variations of V3 than rotational modulation. Radial velocity observations may 
help to decide which possibility is more nearly true because in the first case the 
period would be equal to 0.3084~d, while in the second, it would be twice that long. 
Unfortunately, in both cases the radial-velocity amplitude will be rather small. In the 
case of $\lambda$ Eri-type rotational modulation, temperature induced line-profile 
changes lead to apparent radial-velocity variations with amplitudes of the order of 
10~km~s$^{-1}$ (Balona 1991). In the second case, the star would be an SB1 system 
with the $K_1$ amplitude decreasing with the mass of the secondary component: for the 
secondary's mass equal to 0.8~M$_\odot$ ($\sim$K0), the predicted $K_1$ is equal to 
18~km~s$^{-1}$ for $i$ consistent with the observed $V$ amplitude, and 23~km~s$^{-1}$ 
for $i$ consistent with the $I_{\rm C}$ amplitude; for 0.12~M$_\odot$ ($\sim$M7) these 
numbers become 7~km~s$^{-1}$ and 9~km~s$^{-1}$. 

{\bf V4 = W209.} This star lies outside the Bia{\l }k\'ow field. On the 
single night in 2002, the star was $\sim$0.10~mag brighter in $V$ than on the two 
nights in 2001. 

{\bf V5 = W36.} No spectral classification is available for this star, but in 
the color-magnitude plane (Fig.\ 10) it falls on the cluster's main sequence. Photographic 
$UBV$ photometry of the star was obtained by Hoag {\it et al.\/}~(1961), 
F\"unfschilling (1967), and Moffat and Vogt (1974). Giving equal weights to the 
magnitudes and color indices from these sources, we compute the following mean values: 
$V ={}$12.91$\,\pm\,$0.012~mag, $B-V ={}$0.593$\,\pm\,$0.015~mag, and $U-B 
={}-$0.110$\,\pm\,$0.020 mag. From these numbers, Johnson's (1966) two-color relation 
for luminosity class V stars, and assuming $E(U-B) ={}$0.72$E(B-V)$, we get $E(B-V) 
={}$0.785~mag and $(B-V)_0 ={}-$0.19~mag; the latter value indicates an MK type of 
B4\,V. Finally, we obtain the interstellar-extinction-corrected magnitude, $V_0 = V 
-{}$3.1$E(B-V)$, equal to 10.48~mag. 

A similar exercise performed for V3, using the photoelectric $V$, $B-V$ and $U-B$ from 
Hoag {\it et al.\/}~(1961) as data leads to $E(B-V) ={}$0.635~mag, $(B-V)_0 
={}-$0.175~mag, and $V_0 ={}$10.60~mag. The difference between the last number and 
$V_0$ of V5 is approximately equal to the difference of luminosity class V absolute 
magnitudes estimated from the stars' $(B-V)_0$ indices. Thus, the distance moduli of 
V3 and V5 are approximately equal to each other. Taking into account the fact that V3 
belongs to the cluster (see above), we conclude that V5 is also a member. In addition, 
V5 shows a similar amount of emission at H$\alpha$ as V3 does (see Fig.\ 11). 

%
\begin{figure}[htb] 
\hbox to\hsize{\hss\includegraphics{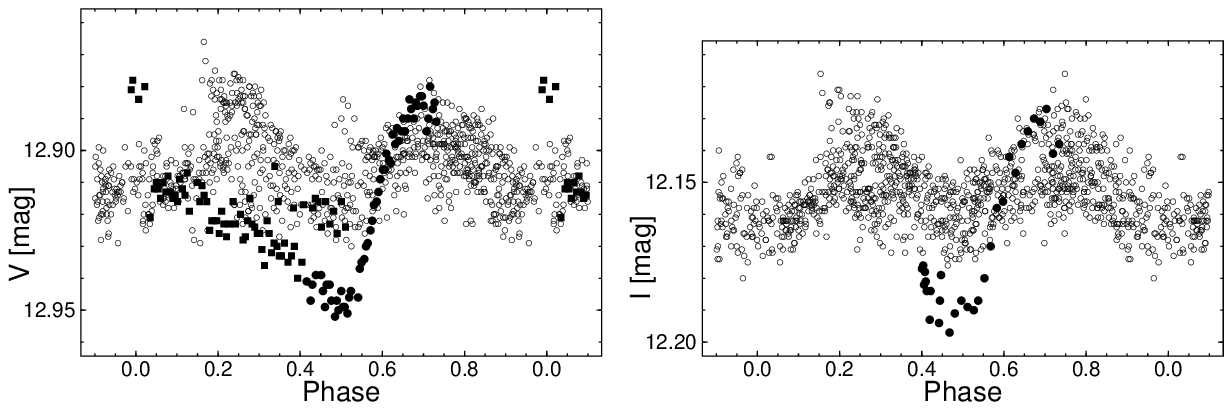}\hss} 
\FigCap{The $V$ (the {\it left-hand panel\/}) and $I_{\rm C}\/$ (the {\it right-hand 
panel\/}) magnitudes of V5 plotted as a function of phase of a period equal to 0.60634~d, 
twice the value derived from the LS spectra. Zero phase corresponds to HJD\,2452110. 
Filled circles are magnitudes from observations taken on JD\,2452102, and squares are 
BOAO magnitudes; there are no squares in the {\it right-hand panel\/} because at BOAO no 
$I_{\rm C}\/$-filter observations of the star were obtained.} 
\end{figure}

The LS power spectra of our $V$ and $I_{\rm C}$ magnitudes of V5 have the highest peaks 
at 3.2985~d$^{-1}$. In Fig.\ 7, the data are plotted as a function of phase of a period 
equal to 0.60634~d, twice the period corresponding to this frequency. The scatter is 
large, but note that it is mainly due to the data obtained on one night, JD\,2452102 
(filled circles in the figure) when the amplitude of the variation was the largest. This 
suggests a periodic variation with variable amplitude, a behavior typical for rotational 
modulation. Assuming the radius of V5 to be equal to 4.2~R$_\odot$ (slightly larger than 
for V3, see the above discussion of the colors and absolute magnitudes of the two stars) 
and the rotation period equal to 0.60634~d, we get the velocity of rotation equal to 
about 60 percent of the brakeup velocity. We conclude that the star is a $\lambda$ 
Eri-type variable with a double-wave light curve. 

{\bf V6 = W55.} This star shows strong H$\alpha$ emission (see Fig.\ 11). We 
have found no light variation in 2001, but on the single night in 2002 the star was 
fainter than in 2001 by $\sim$0.10~mag in $V$, and $\sim$0.06~mag in $I_{\rm C}$. 

{\bf V7 = W79.} No light variation was detected in 2001, but on the single 
night in 2002 the star was fainter than in 2001 by 0.085~mag and 0.045~mag in $V$ and 
$I_{\rm C}$, respectively. 

{\bf V8 = W75.}  The highest peaks in the LS spectra of the $V$ and $I_{\rm C}$ 
magnitudes of the star occur at 1.2052~d$^{-1}$ and 1.2044~d$^{-1}$ for $V$ and $I_{\rm 
C}$, respectively. We adopted a straight mean of these numbers as the correct frequency; 
the corresponding period is equal to 0.83001~d. The amplitudes of a sine-curves of this 
period are equal to 12.9$\,\pm\,$0.6~mmag for $V$ and 11.5$\,\pm\,$0.6~mmag for $I_{\rm 
C}$. The LS spectra of the prewhitened magnitudes show only noise. 

No spectral classification is available for V8, but in the color-magnitude plane 
(Fig.\ 10) the star falls on the cluster's main sequence. Using photoelectric $UBV$ 
photometry of Hoag {\it et al.\/}~(1961), we find that the $E(U-B) ={}$0.72$E(B-V)$ 
reddening line drawn through the star's position in the two-color diagram crosses 
Johnson's (1966) unreddened two-color relation for luminosity class V stars in two 
points. The first point corresponds to $E(B-V) ={}$0.46~mag and $(B-V)_0 ={}$0.18~mag, 
the second, to $E(B-V) ={}$0.65~mag and $(B-V)_0 ={}-$0.01~mag; the first $(B-V)_0$ is 
consistent with MK type $\sim$A6\,V, the second, with $\sim$A0\,V. 

In both these cases the $V$ and $I_{\rm C}$ amplitudes can be accounted for by the 
simple model of ellipsoidal light-variation invoked above in connection with V3. For 
example, for an equal components binary consisting of A6\,V stars and $i ={}$55$^{\rm 
o}$ we get the $V$ and $I_{\rm C}$ amplitudes equal to 13.0~mmag and 11.0~mmag, 
respectively, in a one-$\sigma$ agreement with the observed amplitudes. The same 
result is obtained for an A0\,V primary component and an A5\,V secondary component if 
$i ={}$53$^{\rm o}$. We conclude that V8 is an ellipsoidal variable with the orbital 
period equal to 1.66003~d, twice the value obtained from the LS spectra. The star's 
$V$ and $I_{\rm C}$ magnitudes are plotted as a function of phase of this period in 
the {\em middle left-hand panel\/} of Fig.~6.

{\bf V9 = W62.} No spectral classification is available for this star, but in 
the color-magnitude plane (Fig.\ 10) it falls on the cluster's main sequence, 
indicating luminosity class V. Photographic $UBV$ photometry of V9 was obtained by 
Hoag {\it et al.\/}~(1961), F\"unfschilling (1967), and Moffat and Vogt (1974). 
Giving equal weights to the magnitudes and color indices from these sources, we 
compute the following mean values: $B-V ={}$0.65$\,\pm\,$0.04~mag, and $U-B 
={}$0.20$\,\pm\,$0.05~mag. In the two-color diagram, these numbers place V9 on 
Johnson's (1966) unreddened relation for luminosity class V stars at a position 
corresponding to $\sim$G5\,V. The $E(U-B) ={}$0.72$E(B-V)$ reddening line drawn from 
this position crosses the unreddened relation in two points, (1) and (2).  At (1), 
$E(B-V) ={}$0.25~mag and $(B-V)_0 ={}$0.40~mag, while at (2), $E(B-V) ={}$0.76~mag and 
$(B-V)_0 ={}-$0.10~mag; the first $(B-V)_0$ is consistent with MK type $\sim$F4\,V, the 
second, with $\sim$B8\,V. 

The highest peaks in the LS power spectra of our $V$ and $I_{\rm C}$ magnitudes of V9 
occur at 0.7793~d$^{-1}$. The period corresponding to this frequency is equal to 
1.2832~d. The amplitudes of a sine-curve of this period, fitted to the data by the 
method of least squares, amount to 20.6$\,\pm\,$0.6~mmag for $V$, and 
20.3$\,\pm\,$0.5~mmag for $I_{\rm C}$. The LS spectra of the prewhitened magnitudes 
show only noise. 

In the {\em bottom left-hand panel\/} of Fig.\ 6, the $V$ and $I_{\rm C}$ magnitudes 
are plotted as a function of phase of a period equal to 2.5664~d, twice the value 
derived from the LS spectra. The light variation seen in the figure suggests a binary 
with distorted components. Interestingly, this eliminates the possibility of the 
primary having an MK type of G5\,V or F4\,V because then the system would be so wide 
that the amplitudes predicted by Eq.\ (3) would be much smaller than observed. For 
F4\,V, the maximum amplitudes would be equal to 5~mmag in $V$ and 4~mmag in $I_{\rm 
C}$, and for G5\,V, they would be smaller still, of course. For an equal components 
binary, consisting of B8\,V stars, the predicted amplitudes of the ellipsoidal 
variation are equal to 18.0~mmag and 15.9~mmag in $V$ and $I_{\rm C}$, respectively, 
for $i ={}$67$^{\rm o}$. Thus, reproducing the observed amplitudes requires $i >$ 
67$^{\rm o}$. However, for $i >$ 67$^{\rm o}$, $a \cos i$ is smaller than the sum of 
the component's radii, so that the system would be eclipsing. We conclude that 
explaining the light variation of V9 in terms of binarity requires assuming that the 
system undergoes grazing eclipses. 

In Table 3 we refer to V9 as an EB-type variable, using the term in a somewhat more 
general sense than in the GCVS, where it is restricted to $\beta$ Lyrae-type systems.  

{\bf V10.} This very red star ($V-I_{\rm C} ={}$5.189~mag) lies outside the Bia{\l }k\'ow 
field. From the three BOAO nights we find a year-to-year increase in $V$ amounting to 
$\sim$0.78~mag. 

%
\begin{figure}[htb] 
\hbox to\hsize{\hss\includegraphics{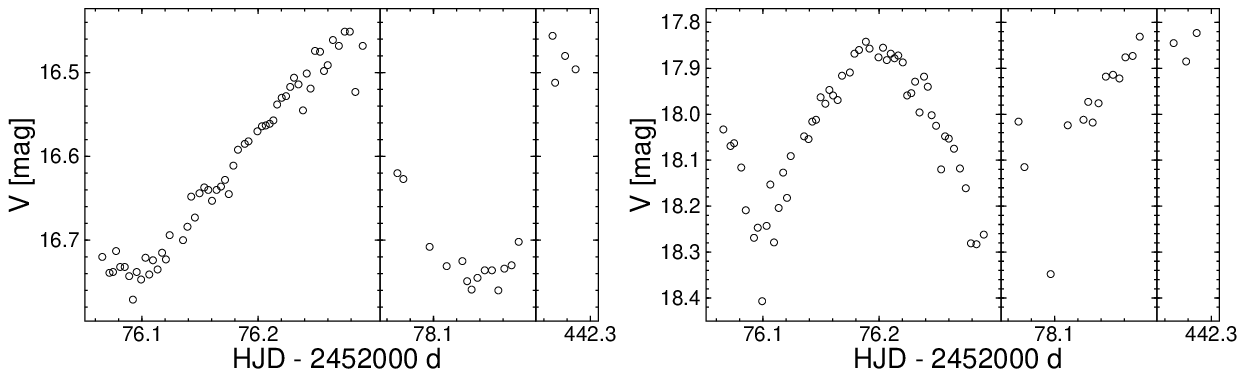}\hss} 
\FigCap{The $V$ magnitudes of V11 (the {\em left-hand panel\/}) and V15 (the {\em 
right-hand panel\/}) plotted as a function of Heliocentric Julian Date.} 
\end{figure}

{\bf V11 = W396.} This star lies outside the Bia{\l }k\'ow field. The BOAO $V$ 
magnitudes are shown in the {\em left-hand panel\/} of Fig.\ 8. The star is probably a W 
UMa-type variable. The 2001 data suggest an orbital period equal to about 0.65~d. 

{\bf V12.} Frequency analysis of our $V$ and $I_{\rm C}$ magnitudes yielded a period of 
1.0065~d. As can be seen from the {\em upper right-hand panel\/} of Fig.\ 6, the star is 
a detached eclipsing binary (EA in GCVS' notation) with the orbital period equal to twice 
this value. The eclipses are partial. They have very nearly the same depth and duration, 
indicating components of similar surface brightness and radius. 

{\bf V13.} Frequency analysis of our $I_{\rm C}$ magnitudes yielded a period of 
0.67401~d. As can be seen from the {\em lower right-hand panel\/} of Fig.\ 6, the star 
is a EB-type system with the orbital period equal to twice this value. 

\begin{table} 
\begin{center} 
\centerline{T a b l e \quad 3} 
{Variable stars in NGC\,6834} 
\vspace{0.3cm} 
{\small 
\begin{tabular} 
{rcccccll} \\ 
\hline\noalign{\smallskip} 
No& WEBDA& $V$&   $V-I_{\rm C}$& $A_V$ or $\Delta V$& $A_I$ or $\Delta I$&\hspace{7pt} $P$& Type\\ 
  &      & [mag]& [mag]&         [mmag]&              [mmag]& \hspace{7pt} [d]& \\ 
\noalign{\smallskip}\hline\noalign{\smallskip}
 V1&  72& 11.596& 0.429& &see Table 2 & & $\gamma$ Dor\cr
 V2&  67& 12.463& 0.950& 230 & 220 &\hspace{10pt} -- & $\gamma$ Cas \cr 
 V3&  23& 12.529& 0.620& 5.2$\pm$0.3& 7.7$\pm$0.3& 0.6168& $\lambda$ Eri? \cr 
 V4& 209& 12.537& 0.774& 103 & -- &\hspace{10pt} -- &unknown \cr
 V5&  36& 12.906& 0.753& 11.2$\pm$0.5 & 8.7$\pm$0.4& 0.60634 & $\lambda$ Eri \cr
 V6&  55& 13.117& 0.841&  98  &  56 &\hspace{10pt} -- &unknown \cr
 V7&  79& 14.296& 0.838&  85  &  45 &\hspace{10pt} -- &unknown \cr
 V8&  75& 14.514& 0.879&12.9$\pm$0.6&11.5$\pm$0.6& 1.66003 & Ell\cr
 V9&  62& 14.740& 0.814&20.6$\pm$0.6&20.3$\pm$0.5& 2.5664& EB\cr 
V10&  --& 16.615& 5.189&  783 & -- &\hspace{10pt} -- &unknown \cr
V11& 396& 16.632& 1.469& 300 & -- & 0.65 & W UMa \cr
V12&  --& 16.635& 1.365& $\sim$36 & $\sim$35 & 2.0130 & EA\cr
V13&  --& 17.567& 2.051& $\sim$23 & $\sim$23 & 1.34802& EB\cr
V14&  --& 17.955& 5.882&  942 & -- &\hspace{10pt} -- &unknown \cr
V15&  --& 17.957& 1.450& 500 & -- & 0.4 & W UMa \cr
\noalign{\smallskip}\hline\noalign{\smallskip}                     
\end{tabular} 
 } 
\end{center} 
\end{table}

{\bf V14.} This star lies outside the Bia{\l }k\'ow field. With $V-I_{\rm C} 
={}$5.882~mag it is redder than V10. The light variation is similar to that of the 
latter star: from the three BOAO nights we find a year-to-year increase in $V$ 
amounting to $\sim$0.94~mag. 

{\bf V15.} This star lies outside the Bia{\l }k\'ow field. The BOAO $V$ magnitudes are 
shown in the {\em right-hand panel\/} of Fig.\ 8. The star is probably a W UMa-type 
variable. From the 2001 data we estimate the orbital period to be equal to about 0.4~d. 

The stars discussed in this section are listed in Table 3. The $V$ magnitudes and 
$V-I_{\rm C}$ color indices in columns three and four are mean values of all data, 
i.e., they were computed from the time-series magnitudes and color indices (see the 
tables in the {\em Acta Astronomica Archive\/}). In case of V12 and V13, only the 
out-of-eclipse data were taken into account in computing the mean values.  Columns 
five and six contain the amplitudes, $A_V$ and $A_I$ with their standard deviations, 
or the ranges, $\Delta V$ and $\Delta I$. The amplitudes were obtained by fitting the 
sine-curves, $A_V \sin (2 \pi t/P_{\rm LS} + \phi_V)$ and $A_I \sin (2 \pi t/P_{\rm 
LS} + \phi_I)$, to the $V$ and $I_{\rm C}$ magnitudes by the method of least squares, 
where $P_{\rm LS}$ is the period derived from the LS spectra. For V12 and V13, 
estimated eclipse depths are given instead of the ranges. The periods, $P$, given in 
column seven are equal to $2P_{\rm LS}$ for most stars; for the W UMa variables V11 
and V15, $P$ is the estimated orbital period, mentioned in the text. The last column 
contains our variability classification. 

\vspace{0.5cm}
\centerline{\bf 4. The color-magnitude plane}

\vspace{0.5cm}
{\em 4.1. The range of color excess}

\vspace{0.5cm}
As we mentioned in the Introduction, Johnson {\it et al.\/}~(1961) found the color 
excess, $E(B-V)$, to vary over the face of NGC\,6834. However, these authors give no 
details, in particular, they do not mention whether they determined the range of 
$E(B-V)$. We shall investigate this point presently using photoelectric $UBV$ data 
from the literature. The result will be used in the next subsection in fitting the 
isochrones to the color-magnitude diagrams.

According to WEBDA, the photoelectric $UBV$ data are available for 28 stars in the 
field of NGC\,6834. Of these, 21 are in the Bia{\l }k\'ow field, 27 in the BOAO field, 
and one is outside the field we observed. For 27 stars, the $V$ magnitudes and the 
$B-V$ and $U-B$ color-indices listed in WEBDA are from Hoag {\em et al.\/}~(1961), for 
one star, W67 (= V2, see Table 3), there are two sets of photoelectric data, one from 
Kozok (1965) and one from Coyne {\em et al.\/}~(1974); in the following we shall use 
mean values of the magnitudes and color indices from these two sources. Numbers 5, 12 
and 14 in Hoag {\em et al.\/}~(1961), to be referred to below as H5, H12, and H14, are 
blends. As already mentioned in Section 3, H5 is a blend of two stars, W72 (= V1) and 
W2098. H12, misidentified in WEBDA as W52, is a blend of two stars, W53 and W2058. 
Finally, H14 = W207 is a blend of W207 and W2054. H12 and H14 will be omitted from the 
following discussion because $E(B-V)$ obtained from blends is a function of the 
unknown color-excesses of the component stars. H5 will be retained because the MK type 
of its brighter component is known, and therefore probable values of the components' 
color-excesses can be estimated (see below). 

All stars can be divided into four groups, depending on how many times their $E(U-B) 
={}$0.72$E(B-V)$ reddening line crosses Johnson's (1966) unreddened relation for 
luminosity class V stars. In group (1) we would include stars with the reddening line 
that crosses Johnson's (1966) relation once, in group (2), those with the reddening 
line crossing the relation twice, in group (3), those having the reddening line cross 
the relation three times, and finally, the stars with the reddening line which misses 
the relation altogether we place in group (4). The situation in which the point 
representing a star falls on the unreddened relation we count as crossing. Stars 
belonging to group (1) have $E(B-V)$ determined unambiguously, those in group (2) have 
two possible values of $E(B-V)$, those in group (3), three values, and for stars in 
group (4) $E(B-V)$ cannot be derived by comparison with the luminosity class V 
unreddened relation. 

The 26 stars in the field of NGC\,6834 with photoelectric $UBV$ photometry, including 
H5 but omitting H12 and H14, are distributed between the four groups as follows: 13 
fall into group (1), seven, into group (2), four, into group (3), and two, into group 
(4). However, in case of two group (2) stars the larger of the two possible values of 
$E(B-V)$ yields intrinsic color-index which is inconsistent with the star's MK type. 
By the same token, the two larger values of $E(B-V)$ of one group (3) star must be 
rejected. Thus, we end up with 16 stars having $E(B-V)$ determined unambiguously, five 
stars with two possible values of $E(B-V)$, three, with three values of $E(B-V)$, and 
two, for which $E(B-V)$ could not be derived by comparison with the luminosity class V 
unreddened relation. 

H5 is one of the two group (2) stars for which we retained the smaller value of 
$E(B-V)$, 0.065~mag in this case. This yields $(B-V)_0 ={}$0.355~mag for the blend, a 
value consistent with the brighter component's MK type of F2\,V, suggesting that 
$E(B-V)$ for the brighter component cannot differ much from 0.065~mag. In order to 
estimate $E(B-V)$ for the fainter component we shall consider the following two cases: 
(1) both stars are reddened by the same amount, equal to 0.065~mag, and (2)~the 
brighter component is reddened by the same amount as in (1), but for the fainter 
component $E(B-V) ={}$0.61~mag, a value equal to the lower bound of $E(B-V)$ for 
NGC\,6834 (see below). Taking the $V$ magnitudes and $V-I_{\rm C}$ color-indices of 
the two stars from Table A1, assuming that the fainter component is a luminosity class 
V star, and that $E(V-I_{\rm C}) ={}$1.25$E(B-V)$ (Dean {\it et al.\/}~1978) and $R_V 
={}$3.1, and using standard relations between color indices (Johnson 1966, Caldwell 
{\em et al.\/}~1993) we de-blended the $UBV$ data and then derived $E(B-V)$ for the 
brighter component. The results were 0.055~mag in case (1), and 0.18~mag in case (2). 
The first value yields $(B-V)_0$ consistent with the MK type, but the second does not. 
We conclude that $E(B-V)$ of the fainter component is smaller than 0.61~mag. In fact, 
it is probably equal to that of the brighter component. 

In addition to V1, the brighter component of H5 just discussed, five further variables, 
viz., V2, V3, V6, V7, and V8, are among the 26 stars for which we derived $E(B-V)$ from 
photoelectric color-indices; for V3 and V8, $E(B-V)$ were obtained already in Section 3. 
In case of V4, V5, V6, and V9, for which photoelectric color-indices are not available, 
we used photographic data from WEBDA; for V5 and V9, $E(B-V)$ were obtained already in 
Section 3. 

All $E(B-V)$ we determined are displayed in Fig.\ 9 as a function of angular distance 
from W4, which we assume to be the center of the cluster. The unambiguously 
determined color excesses are plotted as filled circles, the larger and smaller 
$E(B-V)$ of two possible values are plotted as open circles and triangles, 
respectively, and the maximum, middle and minimum $E(B-V)$ of three possible values 
are plotted as crosses, inverted triangles and open diamonds, respectively. 
Points in open boxes denote $E(B-V)$ obtained from photographic color-indices. The 
point in open box at $\sim$1$^\prime$ corresponds to the largest $E(B-V)$ of V9; the 
two smaller values were rejected in Section 3. In the remaining cases, the photographic 
color-indices yielded single values of $E(B-V)$. The horizontal dashed lines at 
$E(B-V) ={}$0.61~mag and $E(B-V) ={}$0.82~mag were drawn to enclose an area of high 
density of plotted symbols. The vertical dashed line indicates the 3.5$^\prime$ angular 
radius of the cluster (Trumpler 1930). Note that all open circles fall within the 
left-hand area enclosed by the three lines. 

\begin{figure}[htb] 
\hbox to\hsize{\hss\includegraphics{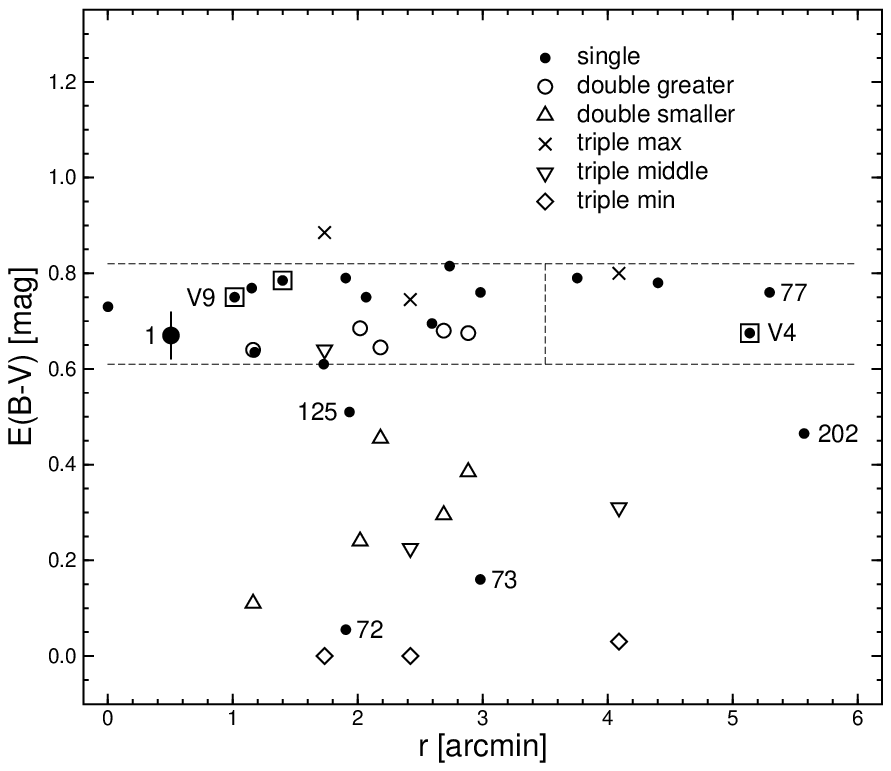}\hss} 
\FigCap{The color excesses, $E(B-V)$, derived from the $UBV$ color-magnitude 
diagram as explained in the text, displayed as a function of angular distance 
from the center of the cluster. The unambiguously determined color excesses are 
plotted as filled circles, the larger and smaller $E(B-V)$ of two possible values are 
plotted as open circles and triangles, respectively, and the maximum, 
middle and minimum $E(B-V)$ of three possible values are plotted as crosses, 
inverted triangles and open diamonds, respectively. The large filled circle with 
error bars represents W1 = HDE\,332843. Points in open boxes denote 
$E(B-V)$ obtained from photographic color-indices. The horizontal dashed lines 
have the ordinates $E(B-V) ={}$0.61~mag and $E(B-V) ={}$0.82~mag. The vertical dashed 
line indicates the 3.5$^\prime$ angular radius of the cluster. Symbols 
representing stars mentioned in the text are labeled with the WEBDA numbers or 
the V numbers from the first column of Table 3.} 
\end{figure}

We shall now examine the cluster membership of the four stars which are represented in 
Fig.\ 9 by filled circles lying outside the horizontal dashed lines. The point labeled 
72 was plotted using $E(B-V)$ of the brighter component, V1 (see above). As we 
mentioned in Section 3, V1 was found to be an F2\,V foreground star by Turner (1976). 
W73 has been classified by Turner (1976) to F8\,V, and likewise found to be a 
foreground star. W125 has been added to the foreground stars by Sowell (1987) on 
the basis of its MK type of G0/5\,III/V. 

The last star with unambiguously determined $E(B-V)$ outside the horizontal dashed 
lines in Fig.\ 9 is W202. The star has been considered to be a member of 
NGC\,6834 by Hoag and Applequist (1965). These authors classified W202 to B5\,IV 
and derived a distance modulus of 10.5~mag. Turner (1976) classified the star to 
B5\,III and determined $M_V ={}-$2.12~mag. From the last number and the 
extinction-corrected $V$ magnitude we get $V_0 - M_V ={}$11.0~mag. These values of 
$V_0 - M_V$ are 0.9~mag to 0.4~mag smaller than the smallest value in the literature 
(see the Introduction). In addition, the angular distance of the star from the center 
of NGC\,6834 is a factor of about two larger than the cluster's diameter (see Fig.\ 9 
and figure 1 of F\"unfschilling 1967). We conclude that the W202 does not belong 
to NGC\,6834. 

In order to estimate how many non-members are enclosed between the horizontal dashed 
lines in Fig.\ 9, let us assume that all symbols outside these lines represent 
non-members. There are 15 such symbols: four filled circles, five triangles, one 
cross, two inverted triangles, and three diamonds. Giving weight 1 to the filled 
circles, 1/2 to the triangles, 1/3 to the cross, inverted open triangles and open 
diamonds, and taking the area outside the horizontal dashed lines to be about three 
times larger that the area within, we get 2.8 as the average number of non-members 
contaminating the area within the horizontal dashed lines. Counting the symbols within 
the horizontal dashed lines in the same way we get 19.5. Thus, about 15 percent of 
stars within the horizontal dashed lines are non-members. Omitting W77 and V4, 
which may be non-members on account of their large angular distance from the center of 
the cluster, would increase this number to 16 percent. In any case, a majority of 
symbols between the horizontal dashed lines in Fig.\ 9 must represent members. Two 
conclusions follow from this, viz., (1) for stars belonging to NGC\,6834, $E(B-V)$ 
ranges from 0.61~mag to 0.82~mag, and (2) because a majority of open circles 
represents members, the majority of triangles should be erased from the figure; this 
decreases the estimate of contamination by non-members to about 10 percent. Note that 
conclusion (2) strengthens conclusion (1). 

Let us now discuss the two group (4) stars, i.e., those for which $E(B-V)$ could 
not be derived by comparison with the luminosity class V unreddened relation. In 
both cases the $E(U-B) ={}$0.72$E(B-V)$ reddening line runs below the unreddened 
relation. The first star, W1 = HDE\,332843, has been classified to F2\,Ib 
by Sowell (1987), and to F0\,Ib by Roslund (1960) and Turner (1976). Thus, the 
unreddened relation to use in this case would be Johnson's (1966) luminosity 
class Ib relation. However, the star's reddening line runs also below this 
relation. The distance between the reddening line and the unreddened relation at 
the ``point of closest approach'' amounts to 0.06~mag. At this point we have 
$E(B-V) ={}$0.67~mag, $(B-V)_0 ={}$0.20~mag, and $(U-B)_0 = (U-B) -{}$0.72$E(B-V) 
={}$0.33~mag. While $(B-V)_0$ is consistent with an MK type of about F0\,Ib 
(note that this has not been assumed), the observed $(U-B)_0$ is about 0.07~mag 
greater than standard $U-B$ for this MK type, but only 0.03~mag greater than 
standard $U-B$ for F2\,Ib. Allowing for the uncertainties in Johnson's (1966) 
unreddened relation for luminosity class Ib stars and errors in the 
photoelectric photometry of Hoag {\it et al.\/}~(1961), we estimate the standard 
deviation of the $E(B-V)$ of W1 = HDE\,332843 to be equal to 0.05~mag. In 
Fig.\ 9, the star is represented by the large filled circle with error bars.

The second group (4) star is W206 = HD\,187971. The star is 10.5$^\prime$ from the 
center of the cluster, outside either field we observed. Hoag and Applequist (1965) have 
classified it to A3\,V. Its photoelectric $B-V$ is consistent with this, but $U-B$ is off 
by 0.050~mag, a discrepancy probably too large to be put down solely to errors in the 
photoelectric photometry of Hoag {\it et al.\/}~(1961). Perhaps the star is peculiar. In 
any case, it is an unreddened or slightly reddened foreground object.  

\vspace{0.5cm}
{\em 4.2. The isochrones}

\vspace{0.5cm}

%
\begin{figure}[htb]
\hbox to\hsize{\hss\includegraphics{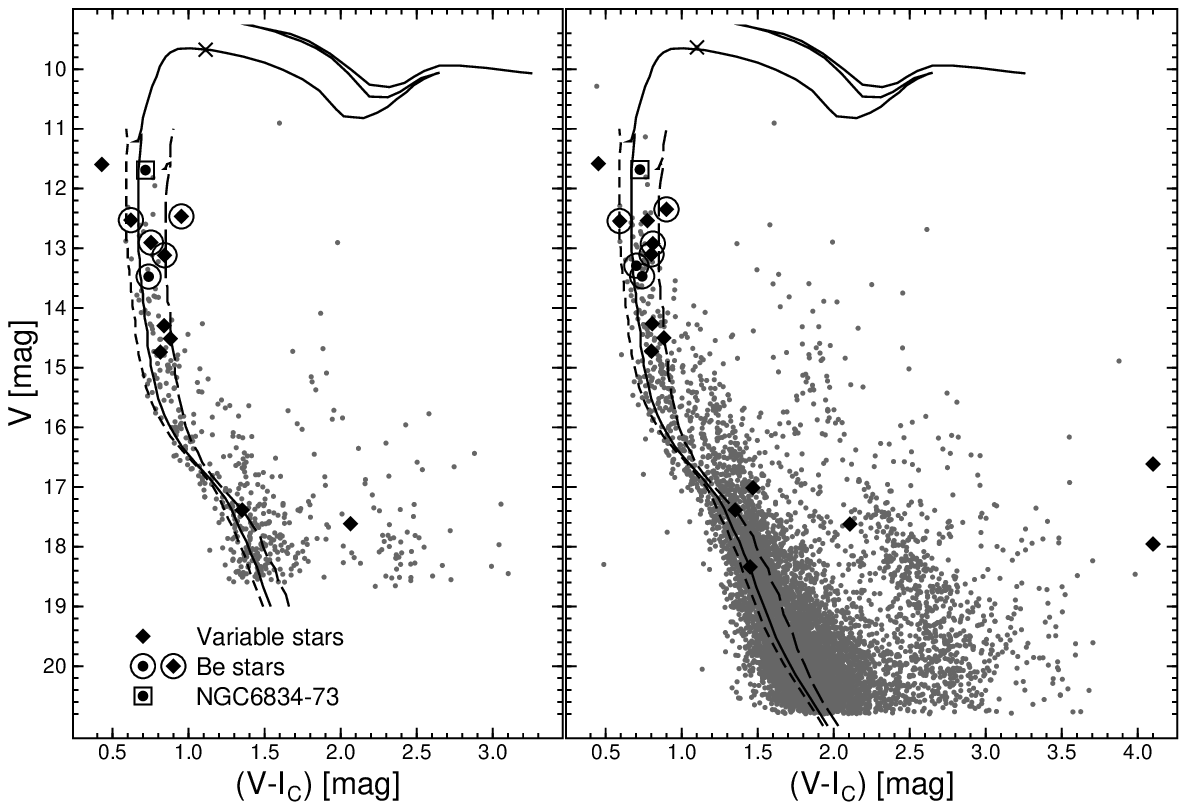}\hss} 
\FigCap{The color--magnitude diagrams for NGC\,6834 from the Bia\l{}k\'ow data 
(the {\em left-hand panel\/}) and from the BOAO data (the {\em right-hand 
panel\/}). Variable stars are shown with diamonds, and the stars with H$\alpha$ 
emission, with symbols in open circles. The two diamonds at far right in the 
{\em right-hand panel\/} are the two very red stars, V10 and V14, which are 
actually redder than shown (see Table 3). The solid circle in an open square is 
W73, the Be star listed in the catalogue of Kohoutek and Wehmeyer (1999) 
for which we have measured a normal $\alpha$ index (see Fig.\ 11). The cross is 
W1 = HDE\,332843, the early-F Ib supergiant. The lines are the $Z 
={}$0.02, $Y ={}$0.28 isochrones from Bertelli {\it et al.\/}~(1994), discussed 
in the text.}
\end{figure}

The $V$ {\em vs.\/} $V-I_{\rm C}$ color--magnitude diagrams for NGC\,6834, based on 
our two data sets, given in Tables A1 and A2, are shown in Fig.~10. The stars we found 
to be variable in light and those showing H$\alpha$ emission (see Section 5) are 
indicated. In case of V11, V12, and V15, duplicity corrections were added to the $V$ 
magnitudes. As can be seen from the figure, the BOAO photometric data are about 2~mag 
deeper in $V$ than the Bia\l{}k\'ow data. Since NGC\,6834 is located close to the 
Galactic plane, the contamination by cluster non-members at fainter magnitudes will be 
much higher than the estimate given in the preceding subsection. 

The lines in Fig.\ 10 are the $Z ={}$0.02, $Y ={}$0.28 isochrones from Bertelli {\it 
et al.\/}~(1994), shifted along the reddening lines $A_V = R_I E(V-I_{\rm C})$ with 
different values of $E(V-I_{\rm C})$; in all cases we have assumed $E(V-I_{\rm 
C})/E(B-V) ={}$1.25 (Dean {\it et al.\/}~1978) and  $R_V ={}$3.1. The solid lines have 
the following parameters: $\log ({\rm age/yr}) ={}$7.70, $V_0 - M_{\rm V} 
={}$12.10~mag, and  $E(V-I_{\rm C}) ={}$0.838~mag, corresponding to $E(B-V) 
={}$0.67~mag, the value determined for W1 = HDE\,332843 in the preceding subsection. 
The short-dashed lines were drawn by shifting the solid line for $V >{}$11~mag to 
$E(V-I_{\rm C}) ={}$0.762~mag, corresponding to $E(B-V) ={}$0.61~mag, the lower bound 
of reddening for NGC\,6834 derived in the preceding subsection. The long-dashed lines 
were drawn by shifting the solid line for $V >{}$11~mag to the upper bound, i.e., 
$E(V-I_{\rm C}) ={}$1.025~mag, corresponding to $E(B-V) ={}$0.82~mag. The 
above-mentioned $\log ({\rm age/yr})$ and $V_0 - M_{\rm V}$ were chosen so that (1) 
the solid lines passed through the cross representing W1 = HDE\,332843, and (2) the 
short-dashed lines formed the left-hand envelope of the plotted points from $V 
\approx{}$11~mag to $V \approx{}$16~mag. In (1) we have assumed, following Sowell 
(1987), that W1 = HDE\,332843 belongs to the cluster. In (2), we take into account the 
fact that below $V \approx{}$16~mag, which on the isochrone shifted to $E(V-I_{\rm C}) 
={}$0.762~mag corresponds to $V_0 - M_{\rm V} \approx{}$12.10~mag, one should expect 
contamination by foreground stars lying to the left of the left-hand envelope. 

The mass of W1 = HDE\,332843 read off the isochrone (the solid lines in Fig.\ 10) is 
equal to 7.0~M$_\odot$. This value is $\sim$30 percent lower than one would associate 
with an early-F Ib MK type. The question whether it is low enough to be inconsistent 
with the star's membership in the cluster is difficult to answer because masses of Ib 
supergiants are poorly known. 

\vspace{0.5cm}
\centerline{\bf 5. Stars with H$\alpha$ emission}

\vspace{0.5cm} 

The $\alpha$ index, derived from the Bia\l{}k\'ow H$\alpha$ observations (see Section 
2), is plotted as a function of the $V$ magnitude and the $V-I_{\rm C}$ color index in 
Fig.~10. As can be seen from the {\em left-hand panel\/} of the figure, four stars, 
W23 = V3, W36 = V5, W55 = V6, and W67 = V2, show strong emission at H$\alpha$, while 
the fifth, W53, exhibits mild emission. As we discussed in Section 4.1, W53 has a close 
optical companion, W2080. The companion accounts for the relatively large standard 
deviation of the star's $\alpha$ index. 

W55 and W67 are among the three Be stars of the cluster included in the 
catalogue of Kohoutek and Wehmeyer (1999). For the third star listed there, W73, 
the $\alpha$ index we measured shows no sign of emission. This is not surprising given 
the transient character of H$\alpha$ emission in Be stars. In the astro-ph preprint of 
their paper, Mathew {\em at al.\/}~(2008) give the coordinates of the four Be stars 
they found in NGC\,6834. These coordinates correspond to W36, W55, 
W67, the strongest H$\alpha$-emission stars in Fig.\ 11, and W106, which 
is outside the field we observed. Unfortunately, no detailed comparison of our results 
can be made with those of Miller {\em et al.\/}~(1996), mentioned in the Introduction, 
because these authors do not specify the stars they found to have emission at 
H$\alpha$. 

%
\begin{figure}[htb]
\hbox to\hsize{\hss\includegraphics{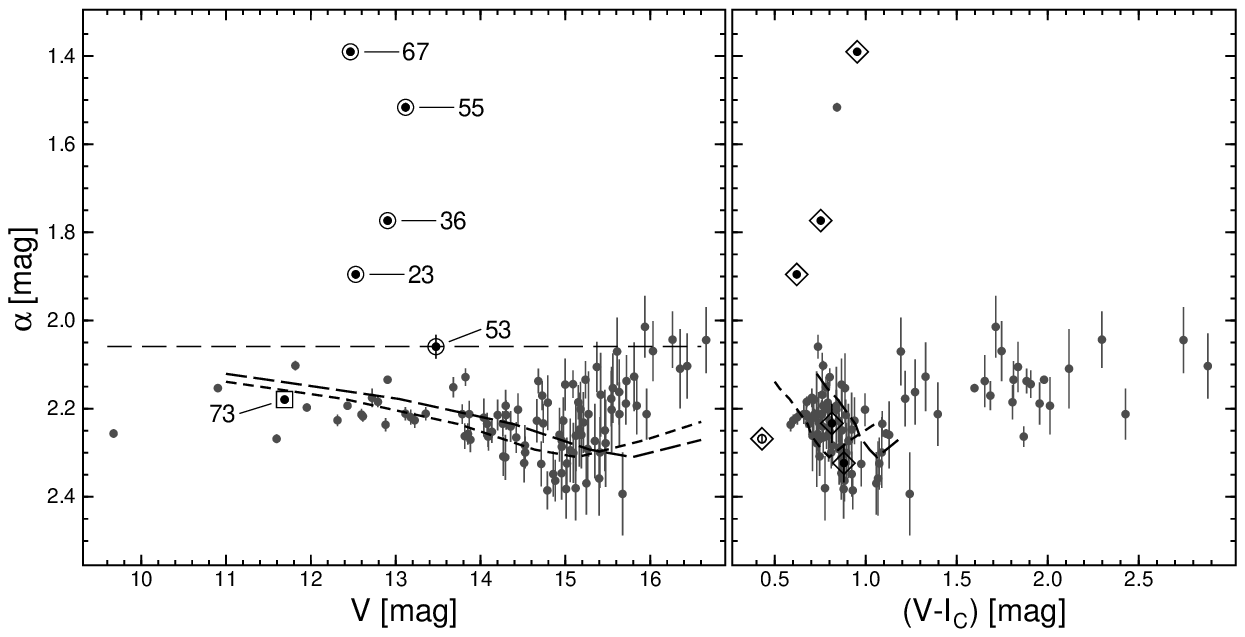}\hss} 
\FigCap{The $\alpha$ index for the brightest stars in the Bia\l{}k\'ow field of 
NGC\,6834 plotted as a function of the $V$ magnitude (the {\em left-hand panel\/}), 
and the $V-I_{\rm C}$ color index (the {\em right-hand panel\/}). In the {\em 
left-hand panel\/}, the stars showing H$\alpha$ emission are shown with points in open 
circles and labeled with their WEBDA numbers. W73 (the point in an open square in the 
{\em left-hand panel\/}), included in the catalogue of Kohoutek and Wehmeyer (1999), 
shows no sign of H$\alpha$ emission. The horizontal long-dashed line indicates the 
expected value of the index for zero equivalent width of the H$\alpha$ line. The 
points in diamonds in the {\em right-hand panel\/} are, from top to bottom, the 
variable stars V2 = W67, V6 = W55, V5 = W36, V3 = W23, V7, V9, V1, and V8. The 
short-dashed lines correspond to the short-dashed lines in Fig.\ 10, while the 
long-dashed lines, to the long-dashed lines in that figure.}
\end{figure}

The short-dashed and long-dashed lines in Fig.\ 11 correspond to the short-dashed and 
long-dashed lines in Fig.\ 10 (terminated at $\sim$17 mag in the case of V and 
$\sim$1.2~mag in the case of $V - I_{\rm C}$), i.e., to the isochrone that fits W1, 
shifted to the lower and upper bound of the color excess. They were computed from the 
$\alpha$-$M_V$ relation for main-sequence stars derived in Paper VI, using the 
parameters mentioned in Section 4.2, i.e., $\log ({\rm age/yr}) ={}$7.70 and $V_0 - 
M_{\rm V} ={}$12.10~mag, and $E(B-V) ={}$0.61~mag and 0.82~mag for the short-dashed 
and long-dashed lines, respectively. 

\vspace{0.5cm}
\centerline{\bf 6. Summary}

\vspace{0.5cm} 
As a result of a CCD variability search in the field of the young open cluster 
NGC\,6834, we discovered 15 stars to be variable in light. The stars, denoted V1, V2, 
..., V15 in order of increasing $V$ magnitude, are discussed in Section 3. For ten of 
them, we determined the type of variability. As can be seen from the last column of 
Table 3, the brightest variable is a $\gamma$ Dor-type star, while the nine 
fainter ones include a $\gamma$ Cas-type variable, two $\lambda$ Eri-type 
stars, an ellipsoidal variable, one EA- and two EB-type eclipsing binaries, and two W 
UMa-type stars. The five variables we could not classify show year-to-year 
variations. 

For 6937 stars, we obtained the $V$ magnitudes and $V-I_{\rm C}$ color-indices on the 
standard system. A direct comparison of these data with theoretical isochrones was 
hampered by the fact that interstellar extinction is variable over the face of the 
cluster (Johnson {\it et al.\/}~1961). In order to circumvent this difficulty, we 
determined $E(B-V)$ for all stars in the field of NGC\,6834 with published 
photoelectric $UBV$ photometry. As demonstrated in Section 4.1, these $E(B-V)$ range 
from $\sim$0.06~mag to $\sim$0.82~mag. However, a majority of the values, including 
that for the early-F Ib supergiant W1 = HDE\,332843, fall into the range from 0.61~mag 
to 0.82~mag. In addition, all stars with unambiguously determined $E(B-V) <{}$0.61~mag 
are non-members. From these two facts we concluded that stars belonging to the cluster 
have $E(B-V)$ bounded by 0.61~mag and 0.82~mag. Then, we fitted properly reddened $Z 
={}$0.02, $Y ={}$0.28 Padova isochrones (Bertelli {\it et al.\/}~1994) to the observed 
$V-I_{\rm C}$, $V$ diagrams in such a way that the left-hand envelope of points 
corresponded to the above-mentioned lower bound of $E(B-V)$. Assuming W1 = HDE\,332843 
to be a member, we got $\log ({\rm age/yr}) ={}$7.70 and $V_0 - M_{\rm V} 
={}$12.10~mag. 

For 103 brightest stars in our field we obtained the $\alpha$ index, a measure of the 
equivalent widths of the H$\alpha$ line. We found H$\alpha$ emission in five stars, 
including the $\gamma$ Cas-type variable, and the two $\lambda$ Eri-type 
variables. For W73, the Be star included in the catalogue of Kohoutek and Wehmeyer 
(1999), we found no sign of H$\alpha$ emission. 

Let us now comment briefly on the cluster membership of the variable stars listed in 
Table 3. As we already mentioned in Section 3, V1, the multiperiodic $\gamma$ 
Dor-type variable, is a foreground star. The remaining eight variable stars 
brighter than $\sim$14.8 V magnitude, i.e., V2, ..., V9, all have $E(B-V)$ falling 
between 0.61~mag and 0.82~mag, the lower and upper bound of $E(B-V)$ for NGC\,6834, 
thus fulfilling the necessary condition to be members. V4 = W209, one of the 
variables we could not classify, is probably a non-member because it lies at an 
angular distance of 5.6$^\prime$ from the center of NGC\,6834, much larger than 
3.5$^\prime$, the cluster's radius (see Fig.\ 9). 

As explained in Section 4.2, the dashed lines in Fig.\ 10 correspond to the lower and 
upper bound of $E(B-V)$ for NGC\,6834. Thus, a star having $E(B-V)$ between these 
bounds should be enclosed within the dashed lines. This is indeed the case for all 
eight variables just discussed but V2 = W67, the $\gamma$ Cas-type star, which falls 
to the right of the long-dashed lines and V6 = W55 which falls on the long-dashed 
lines. Note that the former star shows the strongest H$\alpha$ emission we detected, 
while the latter, the second strongest (see the {\em left-hand panel\/} of Fig.\ 11). 
The explanation of their position with respect to the long-dashed lines is 
straightforward: the two stars exhibit an infrared excess correlated with the strength 
of H$\alpha$ emission, as all other Be stars do (Feinstein and Marraco 1981). 

Of the six variable stars of Table 3 fainter than $\sim$16.5 $V$ magnitude, i.e., V10, 
..., V15, only two, the EA-type eclipsing binary V12 and the W UMa-type variable V15, 
have $V-I_{\rm C}$ and $V$ that place them between the dashed lines in Fig.\ 10. 
Taking this as evidence that their $E(B-V)$ would fall between the lower and upper 
bound for NGC\,6834, we conclude that V12 and V15 could be members. However, V15's 
angular distance from the center of the cluster amounts to 6.0$^\prime$, making its 
membership doubtful. V12, with its angular distance from the center of NGC\,6834 
amounting to 3.0$^\prime$ is more likely to belong to the cluster. The remaining four 
stars are certainly non-members. 

{\bf Acknowledgements.} This work was supported in part by MNiSzW grants NN203 302635 
and NN203 405139. We thank J.\ Molenda-\.Zakowicz for taking part in the observations. 
We have made use of the WEBDA database, operated at the Institute for Astronomy of the 
University of Vienna, the Aladin service, operated at CDS, Strasbourg, France, and the 
SAO/NASA Astrophysics Data System Abstract Service. 

\vspace{0.5cm} 
\centerline{REFERENCES} 
\vspace{0.3cm} {\small 
\refd Balona, L.A.~1990, {\it MNRAS}, {\bf 245}, 92.\par 
\refd Balona, L.A.~1991, in: ``Rapid Variability of OB-stars: Nature and Diagnostic 
Value'', {\em Proc.\ of an ESO Workshop}, Garching bei M\"unchen, Germany, Ed.\ by D.\ 
Baade, p.\ 249.\par 
\refd Becker, W.~1963, {\it Zeitschrift f\"ur Astrophysik}, {\bf 57}, 117.\par 
\refd Bertelli G., Bressan A., Chiosi C., Fagotto F., and Nasi E.~1994, {\it 
Astron.~Astrophys.~Suppl.\ Ser.}, {\bf 106}, 275.\par
\refd Claret, A., D\'{\i}az-Cordov\'es, J., and Gim\'enez, A.~1995, {\it Astron.\ 
Astrophys.\ Suppl.\ Ser.}, {\bf 114}, 247.\par 
\refd Caldwell, J.A.R., Cousins, A.W.J., Ahlers, C.C., van Wamelen, P., and Maritz, 
E.J.~1993, {\it SAAO Circ.}, {\bf 15}, 1.\par
\refd Coyne, G.V., Lee, T.A., and de Graeve, E.~1974, {\em Vatican Obs.\ Publ.} {\bf 1}, 
181.\par
\refd Dean, J.F., Warren, P.R., and Cousins, A.W.J.~1978, {\it MNRAS}, {\bf 183}, 569.\par
\refd D\'{\i}az-Cordov\'es, J., Claret, A., and Gim\'enez, A.~1995, {\it Astron.\ 
Astrophys.\ Suppl.\ Ser.}, {\bf 110}, 329.\par 
\refd Dziembowski, W.A., Daszyñska-Daszkiewicz, J., and Pamyatnykh, A.A.~2007, {\it CoAst}, 
{\bf 150}, 213.\par
\refd Feinstein, A., and Marraco, H.G.~1981, {\it PASP}, {\bf 93}, 110.\par
\refd F\"unfschilling, H.~1967, {\it Zeitschrift f\"ur Astrophysik}, {\bf 66}, 440.\par 
\refd Hoag, A.A., and Applequist, N.L.~1965, {\it Astrophys.~J.~Suppl.~Ser.}, {\bf 12}, 
215.\par 
\refd Hoag, A.A., Johnson, H.L., Iriarte, B., Mitchell, R.I., Hallam, K.L., and 
Sharpless, S.~1961, {\it Publ.~USNO, 2nd Ser.}, Vol.~XVII, Part VII, p.~343.\par 
\refd Jerzykiewicz, M., Pigulski, A., Kopacki, G., Mia{\l }kowska, A., and Niczyporuk, 
S.~1996, {\it Acta Astron.}, {\bf 46}, 253 (Paper I).\par 
\refd Jerzykiewicz, M., Kopacki, G., Molenda-\.Zakowicz, J., and Ko{\l }aczkowski, 
Z.~2003, {\it Acta Astron.}, {\bf 53}, 151 (Paper V).\par 
\refd Johnson, H.L.~1966, {\it ARA\&A}, {\bf 4}, 193.\par
\refd Johnson, H.L., Hoag, A.A., Iriarte, B., Mitchell, R.I., and Hallam, K.L.~1961, {\it 
Lowell Obs.~Bull.}, {\bf 5}, 133.\par 
\refd Kohoutek L., and Wehmeyer R.~1999, {\it Astron.~Astrophys.~Suppl.~Ser.}, {\bf 
134}, 255.\par 
\refd Ko{\l }aczkowski, Z., Pigulski, A., Kopacki, G., and Michalska, G.~2004, {\it 
Acta Astron.}, {\bf 54}, 33 (Paper VI).\par 
\refd Kopacki, G., Drobek, D., Ko{\l }aczkowski, Z., and Po{\l }ubek, G.~2008, {\it 
Acta Astron.}, {\bf 58}, 373.\par 
\refd Kozok, J.K.~1985, {\it Astron.\ Astrophys.\ Suppl.\ Ser.}, {\bf 61}, 387.\par 
\refd Lang, K.R.~1992, {\em Astrophysical Data}, Springer.\par 
\refd Landolt, A.U.~1992, {\it Astron.\ J.}, {\bf 104}, 340.\par 
\refd Mathew, B., Subramaniam, A., and Bhatt, B.C.~2008, {\it MNRAS}, {\bf 388}, 
1879.\par 
\refd Michalska, G., Pigulski, A., St\c{e}\'slicki, M., and Narwid, A.~2009, {\it Acta 
Astron.}, {\bf 59}, 349 (Paper VII).\par 
\refd Miller, G.J., Grebel, E.K., and Yoss, K.M.~1996, {\it Bull.\ AAS}, {\bf 28}, 
1367.\par 
\refd Moffat, A.F.J.~1972, {\it Astron.~Astrophys.~Suppl.~Ser.}, {\bf 7}, 355.\par 
\refd Moffat, A.F.J., and Vogt, N.~1974, {\em Veroeff.\ Astron.\ Inst.\ Ruhr-Univ.\ 
Bochum}, {\bf 2}, 1.\par 
\refd Morris, S.L.~1985, {\it Astrophys.~J.}, {\bf 295}, 143.\par 
\refd Paunzen, E., Netopil, M., Iliev, I.Kh., Maitzen, H.M., Claret, A., and Pintado, 
O.I.~2006, {\it Astron.~Astrophys.}, {\bf 454}, 171.\par 
\refd Pigulski, A., Jerzykiewicz, M., and Kopacki, G.~1997, {\it Acta Astron.}, {\bf 
47}, 365 (Paper II).\par 
\refd Pigulski, A., Ko{\l }aczkowski, Z., and Kopacki, G.~2000, {\it Acta Astron.}, 
{\bf 50}, 113 (Paper III).\par 
\refd Pigulski, A., Kopacki, G., and Ko{\l }aczkowski, Z.~2001, {\it Acta Astron.}, 
{\bf 51}, 159 (Paper IV).\par 
\refd Roslund, C.~1960, {\it PASP}, {\bf 72}, 205.\par 
\refd Ruci\'nski, S.M.~1969, {\it Acta Astron.}, {\bf 19}, 125.\par 
\refd Ruprecht, J.~1966, {\it Bull.~Astron.~Czech.}, {\bf 17}, 33.\par 
\refd Russell, H.N., and Merrill, J.E.~1952, {\it Contrib.\ Princeton Univ.\ Obs.}, 
No.~26.\par 
\refd Slettebak, A.~1966, {\it Astrophys.~J.}, {\bf 145}, 126.\par 
\refd Sowell, J.R.~1987,  {\it Astrophys.~J.~Suppl.~Ser.}, {\bf 64}, 241.\par 
\refd Stetson, P.B.~1987, {\it PASP}, {\bf 99}, 191.\par
\refd Stetson, P.B.~1990, {\it PASP}, {\bf 102}, 932.\par
\refd Trumpler, R.J.~1930, {\it Lick Obs.~Bull.}, {\bf 14}, 154.\par 
\refd Turner, D.G.~1976, {\it Astron.~J.}, {\bf 81}, 1125.\par 
\refd Voroshilov, V.I., Kalandadze, N.B., Kolesnik, L.N., Polishchuk, E.P., and 
Fedorochenko, G.L.~1969, {\it Catalogue of B- and V-magnitudes of 12000 stars}, Kiev: 
Naukova dumka.\par
\refd Walker, G.A.H., Kuschnig, R., Matthews, J.M., Cameron, C., Saio, H., Lee, U., 
Kambe, E., Masuda, S., Guenther, D.B., Moffat, A.F.J., Rucinski, S.M., Sasselov, D., 
and Weiss, W.W.~2005, {\it ApJ}, {\bf 635}, L77.\par
}
\end{document}